\documentclass[aps,prd,showpacs,amsfonts,notitlepage,amssymb,amsmath,nofootinbib,superscriptaddress]{revtex4-1}
\usepackage[utf8]{inputenc}
\usepackage[english]{babel}
\usepackage[titletoc,toc,title]{appendix}
\usepackage{amsmath}
\usepackage{amsfonts}
\usepackage[colorlinks=true,linkcolor=blue,citecolor=blue,urlcolor=blue]{hyperref}
\usepackage{amssymb}
\usepackage{graphicx}
\usepackage{enumerate}
\usepackage{csquotes}
\usepackage[caption=false]{subfig}
\usepackage{dsfont}
\usepackage{color}
\usepackage{bbold}
\usepackage{mathtools}
\usepackage{tensor}

\newcommand{\eq}[1]{Eq.~(\ref{#1})}

\newcommand{\dd}{\ensuremath{\mathrm{d}}}

\begin{document}

\title{Assessing degrees of entanglement of phonon states in atomic Bose gases \\
through the measurement of commuting observables}  
\date{\today}

\author{Scott Robertson\thanks{scott.robertson@th.u-psud.fr}}
\affiliation{Laboratoire de Physique Th\'eorique (UMR 8627), CNRS, Univ. Paris-Sud, Universit\'e Paris-Saclay, 91405 Orsay, France}
\author{Florent Michel\thanks{florent.michel@th.u-psud.fr}}  
\affiliation{Laboratoire de Physique Th\'eorique (UMR 8627), CNRS, Univ. Paris-Sud, Universit\'e Paris-Saclay, 91405 Orsay, France}
\affiliation{Center for Particle Theory, Durham University, South Road, Durham, DHA 3LE, UK} 
\author{Renaud Parentani\thanks{renaud.parentani@th.u-psud.fr}}
\affiliation{Laboratoire de Physique Th\'eorique (UMR 8627), CNRS, Univ. Paris-Sud, Universit\'e Paris-Saclay, 91405 Orsay, France}

\begin{abstract}

We show that measuring commuting observables can be sufficient to assess that a 
bipartite 
state 
is entangled according to either nonseparability or the stronger criterion of `steerability'. 
Indeed, the measurement of a single observable might reveal the strength of the interferences between the two subsystems, 
as if an interferometer were used. For definiteness we focus on the two-point correlation function of density fluctuations 
obtained by {\it in situ} measurements in homogeneous one-dimensional cold atomic Bose gases.
We then compare this situation to that found in transonic stationary flows mimicking a black hole geometry
where correlated phonon pairs are emitted on either side of the sonic horizon by the analogue Hawking effect. 
We briefly apply our considerations to two recent experiments.  

\end{abstract}

\maketitle

\section{Introduction}

Quantum field theory allows the creation of pairs of correlated (quasi-)particles
via strong variations of the classical background~\cite{Birrell-Davies}. 
When focusing on the correlations between the two particles, two cases are particularly clear. 
Firstly, when the background is homogeneous and time-dependent, one obtains 
pairs of quanta with opposite wavenumbers, as is the case 
in an expanding homogeneous universe; 
see Refs.~\cite{Carusotto-DCE,Jaskula-et-al,Robertson-Michel-Parentani-2017,Campo-Parentani-2004} for works discussing these correlations. 
Secondly, 
when the background is stationary but inhomogeneous the pairs of created quanta 
carry opposite energies, as is the case for electro-production in a constant electric field 
and for the (Hawking) radiation emitted by a black hole~\cite{Primer}. 

In both cases, pair production can be stimulated by quasi-particles already present, or emerge via excitation of vacuum fluctuations. 
The latter contribution is of particular interest, as it gives rise to entangled 
states. 
Entanglement is a well-defined notion for pure states; 
for mixed states, on the other hand, 
some care is needed to properly define which subset is to be considered `entangled'.  
Indeed, 
historically, several 
inequivalent 
criteria have been discussed and compared, see e.g.~\cite{Jones-Wiseman-Doherty-PRA}.
In this paper we only consider two of them: nonseparability~\cite{Werner,Simon}, which is particularly simple,
and the older and stronger criterion based on the possibility of steering the outcome of a measurement on a subsystem 
having already measured the state of its partner~\cite{Wiseman-Jones-Doherty-PRL}. 
These notions are recalled in Appendix~\ref{app:Entanglement} 
for bosonic degrees of freedom, 
which is the case we shall consider in this paper. 
For each of these criteria, we also present some inequalities relating observable quantities 
that are sufficient for the criterion to be satisfied. This step is crucial as it 
translates the 
criterion, which 
is defined 
in rather abstract terms, at the level of observables. 
For instance, in homogeneous systems, 
the bipartite state of phonons of wavevectors $k, -k$ is necessarily nonseparable whenever the following inequality is 
satisfied~\cite{Campo-Parentani-2005,Adamek-Busch-Parentani,deNova-Sols-Zapata}:
\begin{equation}
n_k\, n_{-k} - |c_k|^2 < 0 \, , 
\label{nonsep}
\end{equation}
where $n_{\pm k} = \langle \hat b^\dagger_{\pm k} \hat b_{\pm k} \rangle$ give the mean occupation number of particles with wavenumber $\pm k$, and the norm of $c_k = \langle \hat b_{ k} \hat b_{ -k} \rangle$ accounts for the strength of the 
correlation between the $k$ and $-k$ quanta.

Having identified the relevant 
inequalities, one should then address the question of their observability, that is, identify possible sets 
of measurements which are sufficient to assess that the inequality is violated, and therefore that the state under consideration 
is necessarily entangled. At first sight, it seems natural to consider measurements of non-commuting 
observables. In fact, 
to be able to verify that some Bell inequality is violated, it is necessary to consider some set of non-commuting 
observables,
 see~\cite{Banaszek-1998,Campo-Parentani-2006} for bosonic degrees of freedom. 
This is the line of thought that was adopted in~\cite{Finazzi-Carusotto-2014} 
and further advocated in~\cite{Finke-Jain-Weinfurtner}. However, when considering pair creation of quasi-particles of opposite wave number $k,-k$ in homogeneous systems, it was noticed in~\cite{Busch-Parentani-2013,Busch-Carusotto-Parentani} and further clarified 
in~\cite{Robertson-Michel-Parentani-2017} that
{\it in situ} measurements of 
the $k$-th Fourier component of the connected part of the density-density correlation function at some time $t$, $G^{(2)}(k;t)$, 
can be sufficient to assess the nonseparability of the state. The reason is that 
the value of $G^{(2)}$, viewed as a function of time, periodically oscillates due to
interferences between the $k$ and $-k$ sectors, the entanglement of which being the addressed question. As a result, 
when the minimal value of $G^{(2)}(k;t)$ goes 
below $G^{(2)}_{\rm vac}(k)$ evaluated in the phonon vacuum state, this guarantees that inequality~(\ref{nonsep}) is satisfied 
which in turn implies that the bipartite state $k,-k$ 
is nonseparable. 

In the present paper we pursue the analysis undertaken in~\cite{Robertson-Michel-Parentani-2017}.
We clarify and extend it in several directions, first by considering the stronger criterion of ``steerability'', then by distinguishing the isotropic and anisotropic homogeneous cases. 
In this paper we also consider globally {\it inhomogeneous} 
background flows which contain two homogeneous domains where the two-point function $G^{(2)}$ can be analyzed in $k$-space.
Our motivation there 
is to reconsider the entanglement of phonon pairs produced in a stationary transonic flow by the Hawking effect of an analogue black hole following the observability~\cite{Steinhauer-2015,deNova-Sols-Zapata}, and the experimental implementation~\cite{Steinhauer-2016}, of the criterion studied in~\cite{Busch-Parentani-2014}. 

The paper is organized as follows. In Section~\ref{sec:Homogeneous}, we study density fluctuations in globally homogeneous backgrounds. We demonstrate that {\it in situ} measurements of density fluctuations performed at a given time 
can contain enough information to assess that the bipartite phonon state of wave numbers $k,-k$ is nonseparable,
or even obeys the stronger criterion of steerability. In Section~\ref{sec:Inhomogeneous}, we study the statistical properties of the 
density fluctuations in the homogeneous domains of a stationary transonic flow mimicking a black hole geometry.  Here
the 
state under study contains two phonon modes carrying opposite energy which propagate against the background flow, and 
a third phonon mode which is co-propagating with respect to the flow and hardly contributes to the relevant expectation values. 
However, we show that the latter mode 
plays an essential 
role as it ensures 
that the measured quantities commute with each other. We conclude in Section~\ref{sec:Conclusion}. In Appendix~\ref{app:Entanglement}, we recall the two notions of entanglement described above, 
while in Appendix~\ref{app:OtherMeasurements} we consider the extra information about the phonon state one could extract 
by measuring phase fluctuations (which do not commute with measurements of the density).

\section{{\it In situ} measurements of atomic density fluctuations
\label{sec:Homogeneous}} 

For simplicity and definiteness, we consider elongated (i.e., effectively one-dimensional) atomic condensates, 
with transverse dimensions much smaller than their length~\cite{Menotti-Stringari,Tozzo-Dalfovo,Gerbier}. 
Quasi-particle excitations of longitudinal momentum $k$ and energy $\omega_{k}$ (we work in units where $\hbar \equiv 1$) are 
well-defined when the background condensate is homogeneous and stationary in a sufficiently large domain
with respect to $1/k$ and $1/\omega_{k}$. For simplicity, we shall thus assume 
local homogeneity and stationarity of the background when measuring density fluctuations, 
though in general the background will be globally inhomogeneous and/or nonstationary.  
In Sec.~\ref{sub:Generalities} we introduce the relevant quantities applicable to the most general settings, while in Secs.~\ref{sub:Homogeneous} and~\ref{sub:HomoEntanglement} we restrict our attention to globablly homogeneous systems. 
The extension to inhomogeneous time-independent systems is delayed until Sec.~\ref{sec:Inhomogeneous}.  

\subsection{Generalities
\label{sub:Generalities}}  

We work in the standard second-quantized formalism 
and adopt the Bogoliubov approximation~\cite{Dalfovo-et-al-1999,Pitaevskii-Stringari-BEC}, where 
the field operator for the 
dilute Bose gas is written 
$\hat{\Phi}(t,x) = e^{-i \mu t + i K x} \left( \Phi_{0}
 + \delta\hat{\phi}(t,x) \right)$, 
where $\Phi_{0}$ is a $c$-number that describes the condensed fraction of the gas, 
$\mu$ is the chemical potential, 
$K$ is the condensate momentum, 
and $\delta\hat{\phi}$ 
describes 
perturbations on top of the condensate.  
Since we assume 
(local)  
homogeneity and stationarity of the background, $\left|\Phi_{0}\right|^{2} \equiv \rho_{0}$ is constant, and is equal to the one-dimensional number density of condensed atoms.  We can also refer to these as atoms with zero momentum (relative to the condensate).
Since we have explicitly factored out the spatial component of the condensate wave function, 
$\Phi_{0}$ is independent of $x$, and can be taken to be real and positive so that $\Phi_{0} = \sqrt{\rho_{0}}$. 
As an operator, the total atom number density is
\begin{eqnarray}
\hat{\rho}(t,x) & = & \hat{\Phi}^{\dagger}(t,x) \hat{\Phi}(t,x)\, , \nonumber \\
& \approx & \rho_{0} + \sqrt{\rho_{0}} \left(\delta\hat{\phi}(t,x) + \delta\hat{\phi}^{\dagger}(t,x) \right) \,, 
\label{density}
\end{eqnarray}
where in the second line we have neglected the nonlinear contribution of the perturbations.  
Linear density fluctuations are thus described by the operator $\delta\hat \rho = 
\sqrt{\rho_{0}} \left( \delta\hat{\phi} + \delta\hat{\phi}^{\dagger} \right) $. 
In the body of this paper, we shall only use {\it in situ} measurements of $\delta\hat \rho(t,x)$
performed at some time $t$. 
Using the equal-time commutators $[\hat{\Phi}(t,x), \hat{\Phi}(t,x')] = 0$ and 
$[\hat{\Phi}(t,x), \hat{\Phi}^\dagger(t,x')] = \delta(x-x')$, 
one easily verifies that $\delta\hat \rho(t,x)$ and $\delta\hat \rho(t,x')$   
commute with each other. Hence only commuting measurements are considered in what follows.

At quadratic order, the statistical properties of $\delta\hat \rho(t,x)$ are encoded in the (equal-time) two-point function 
\begin{eqnarray}
G^{(2)}(t,x; t,x') &=& \left\langle \delta\hat \rho(t,x) \, \delta\hat \rho(t,x') \right\rangle \, , \nonumber \\
&=& \left\langle \hat \rho(t,x) \, \hat \rho(t,x') \right\rangle - \left\langle \hat \rho(t,x)\right\rangle \left\langle \hat \rho(t,x') \right\rangle \, ,
\label{G2}
\end{eqnarray}
where the expression on the second line makes clear that $G^{(2)}(t,x; t,x')$ is the connected part of the density-density
correlation function. Hence the contributions of coherent states of phonons are 
removed by the 
subtraction. 
This two-point 
function has been experimentally studied by repeated measurements of $\rho(x)$ 
in Refs.~\cite{Schley-et-al,Steinhauer-2014,Steinhauer-2016}, which motivated the present study.~\footnote{To see what additional information can be extracted from phase measurements, 
in Appendix~\ref{app:OtherMeasurements} we study other two-point functions involving the phase fluctuation $\delta \hat{\theta}$, 
which does not commute with $\delta\hat{\rho}$, even at equal time.}  

The density fluctuations include all atoms carrying a non-zero momentum (relative to the condensate), and thus all non-constant Fourier components of the density profile.  Indeed, it is useful to invoke homogeneity of the background to write explicitly the Fourier components of the density fluctuations
and the density-density two-point function: 
for a region of length $L$ and a wave vector $k \in 2\pi \mathbb{Z} /L$, we have 
\begin{equation}
G^{(2)}(t,k; t,k') \equiv \int_{0}^{L} \dd x \, e^{-i k x} \int_{0}^{L} \dd x' \, e^{i k' x'} G^{(2)}(t,x; t,x') = \left\langle \hat{\rho}_{k}(t) \hat{\rho}_{k'}^{\dagger}(t) \right\rangle \,,  
\label{G2Fourier}
\end{equation}
where
\begin{equation}
\hat{\rho}_{k}(t) \equiv \int_{0}^{L} e^{-ik x}\hat{\rho}(t,x) \dd x \,. 
\label{FTdefn}
\end{equation}
Because $\hat{\rho}(t,x)$ is a hermitian operator, 
it follows immediately from~(\ref{FTdefn}) that 
$\hat{\rho}_{k}^{\dagger}(t) = \hat{\rho}_{-k}(t)$,
and therefore (since at equal time different Fourier components $\hat{\rho}_{k}$ always commute) that 
$\hat{\rho}_{k}(t)$ and $\hat{\rho}_{k}^{\dagger}(t)$ commute with each other; 
the ordering of the operators on the right-hand side of Eq.~(\ref{G2Fourier}) is thus irrelevant.  
Returning to Eq.~(\ref{density}) for the expression for the density fluctuations, we get 
\begin{equation}  \hat{\rho}_{k}(t) = \delta \hat{\rho}_{k}(t) =  \sqrt{N} 
\left( \hat{\phi}_{k}(t) + \hat{\phi}_{-k}^{\dagger}(t) \right)\,, 
\label{densityFT}
\end{equation}
where $N= \rho_0 L$ is the number of condensed atoms in the region of length $L$, and where the 
normalization factor 
$\sqrt{N}$ 
has been chosen so that the operators $\hat{\phi}_{k}$ satisfy the standard equal-time commutation relation
\begin{equation}
\left[ \hat{\phi}_{k}(t) \,, \hat{\phi}_{k^{\prime}}^{\dagger}(t) \right] = \delta_{k,k^{\prime}} \,.
\label{commutator}
\end{equation}
Note the two contributions to $\hat{\rho}_{k}$ in \eq{densityFT}, 
the first of which destroys an atom of momentum $k$, and the second of 
which creates an atom of momentum $-k$.  
The appearance of these two operators is required for the identity $\hat{\rho}_{k}^{\dagger}(t) = \hat{\rho}_{-k}(t)$ to be satisfied, and is thus instrumental in ensuring that $\hat{\rho}_{k}(t)$ commutes with $\hat{\rho}_{k}^{\dagger}(t)$.  
Both 
have the effect of reducing the momentum by $k$, and are indistinguishable 
when measuring the atomic density at a given time. 
As a result they interfere when evaluating the Fourier components 
of the two-point function 
in Eq.~(\ref{G2Fourier}). 
It is precisely these interferences that we shall exploit for assessing the nonseparability of the state. 

%
%
It should also be noticed that measurements of $\hat{\rho}_{k}(t)$ 
performed 
at different times do not generally commute. 
When we refer to such measurements, we do so in a 
``weakly non-commuting'' sense: the measurements would be 
non-commuting if performed on the same experimental realization, 
but 
(since the condensate is generally destroyed when the density profile
$\rho(x)$ is measured~\footnote{This is not necessarily the case if one follows the rather sophisticated method proposed in~\cite{Finazzi-Carusotto-2014}.}) 
the measurements are actually performed on different realizations of the same system. 

As explained in textbooks~\cite{Pitaevskii-Stringari-BEC,Pethick-Smith-BEC}, the Hamiltonian of linear perturbations 
is diagonalized by writing $\hat{\phi}_{k} = u_{k} \hat{\varphi}_{k} + v_{k} \hat{\varphi}_{-k}^{\dagger}$, where $u_{k}^{2}-v_{k}^{2}=1$~\footnote{$u_{k}$ and $v_{k}$ are then uniquely defined by the additional relation $u_{k}/v_{k} = -\left( 1 + k^{2}\xi^{2}/2 + k\xi \sqrt{1+k^{2}\xi^{2}/4} \right)$, where $\xi$ is the healing length defined later in the paragraph.}  
in order to preserve the bosonic commutation relation (\ref{commutator}) with $\hat \phi_k$ replaced by $\hat \varphi_k$. 
Then \eq{densityFT} is equivalent to 
\begin{equation}
\hat{\rho}_{k}(t) = \sqrt{N} 
\left( u_{k}+v_{k} \right) \left( \hat{\varphi}_{k}(t) + \hat{\varphi}_{-k}^{\dagger}(t) \right) \,.
\label{densityFTphonon}
\end{equation}
The operators $\hat{\varphi}_{k}$ correspond to collective excitations (phonons), with $\hat{\varphi}_{k}$ ($\hat{\varphi}_{k}^{\dagger}$) destroying (creating) a phonon of momentum $k$ relative to the condensate.  The fact that they diagonalize the Hamiltonian means that (so long as the background is stationary) the $k$ and $-k$ sectors decouple, so that we can write
\begin{alignat}{2}
\hat{\varphi}_{k}(t) = \hat{b}_{k} \, e^{-i \omega_{k} t}\,, & \qquad \hat{\varphi}_{-k}^{\dagger}(t) = \hat{b}_{-k}^{\dagger} \, e^{i \omega_{-k} t} \,, 
\label{phonon_operators}
\end{alignat}
where the operators $\hat{b}_{k}$ and $\hat{b}_{-k}^{\dagger}$ do not depend on time.
The lab frame frequency $\omega_{k}$ is related to the condensate rest frame frequency $\Omega_{k}$ by a Doppler shift: 
\begin{equation}
\omega_k-vk= \Omega_{k} \equiv c \left|k\right| \sqrt{1 + k^2\xi^{2}/4} \, ,  
\label{disprel} \end{equation} where 
$c = \sqrt{g \rho_{0}/m}$ is the velocity of the low-frequency phonons, 
$\xi = 1/mc$ is the healing length, 
and $v=K/m$ is the flow velocity of the condensate. 
Here $\Omega_{k}$ is taken to be positive, since the negative-frequency solutions are automatically accounted for by the hermitian conjugate operators in the second of Eqs.~(\ref{phonon_operators}). 

\subsection{Homogeneous systems
\label{sub:Homogeneous}} 

Using these observables, 
let us first analyze 
a condensate which is globally homogeneous, and a phonon state which is statistically homogeneous; 
that is, whose two-point function depends only on the spatial interval $x-x'$, and not on $x$ and $x'$ individually.  
Then the expectation value of $\hat{\rho}_{k}(t) \hat{\rho}_{k'}^\dagger(t)$ can be non-zero only when $k=k'$;  
the non-trivial part of the two-point function is therefore simply the autocorrelation $\left\langle \hat{\rho}_{k}(t) \hat{\rho}_{k}^{\dagger}(t) \right\rangle$, and can be expressed in terms of only $k$ and $t$. 
As a result, at quadratic order in the phonon operators, the 
relevant quantity is 
\begin{eqnarray}
G^{(2)}(k,t) & = & N \,  \left( u_{k} + v_{k} \right)^{2} \left( 1 + n_{k} + n_{-k} + 2 \,\mathrm{Re}\left[c_{k} \, e^{-2i\Omega_{k} 
t}\right] \right) \,.
\label{density-density}
\end{eqnarray}
which 
is fully governed by the expectation values 
\begin{alignat}{2}
n_{\pm k} = \left\langle \hat{b}^{\dagger}_{\pm k} \hat{b}_{\pm k} \right\rangle \,, & 
\qquad c_{k} = \left\langle \hat{b}_{k} \hat{b}_{-k} \right\rangle \,.
\label{n_c_defn}
\end{alignat}
The mean occupation numbers $n_{\pm k}$ 
are real and positive, while $c_k$ is in general a complex number. 
Whenever the background is stationary, $n_{\pm k}$ and $c_k$ are constant in time.  
Note that, being a function of $k$ only, Eq.~(\ref{density-density}) is manifestly Galilean invariant. 

\begin{figure}
\includegraphics[width=0.45\columnwidth]{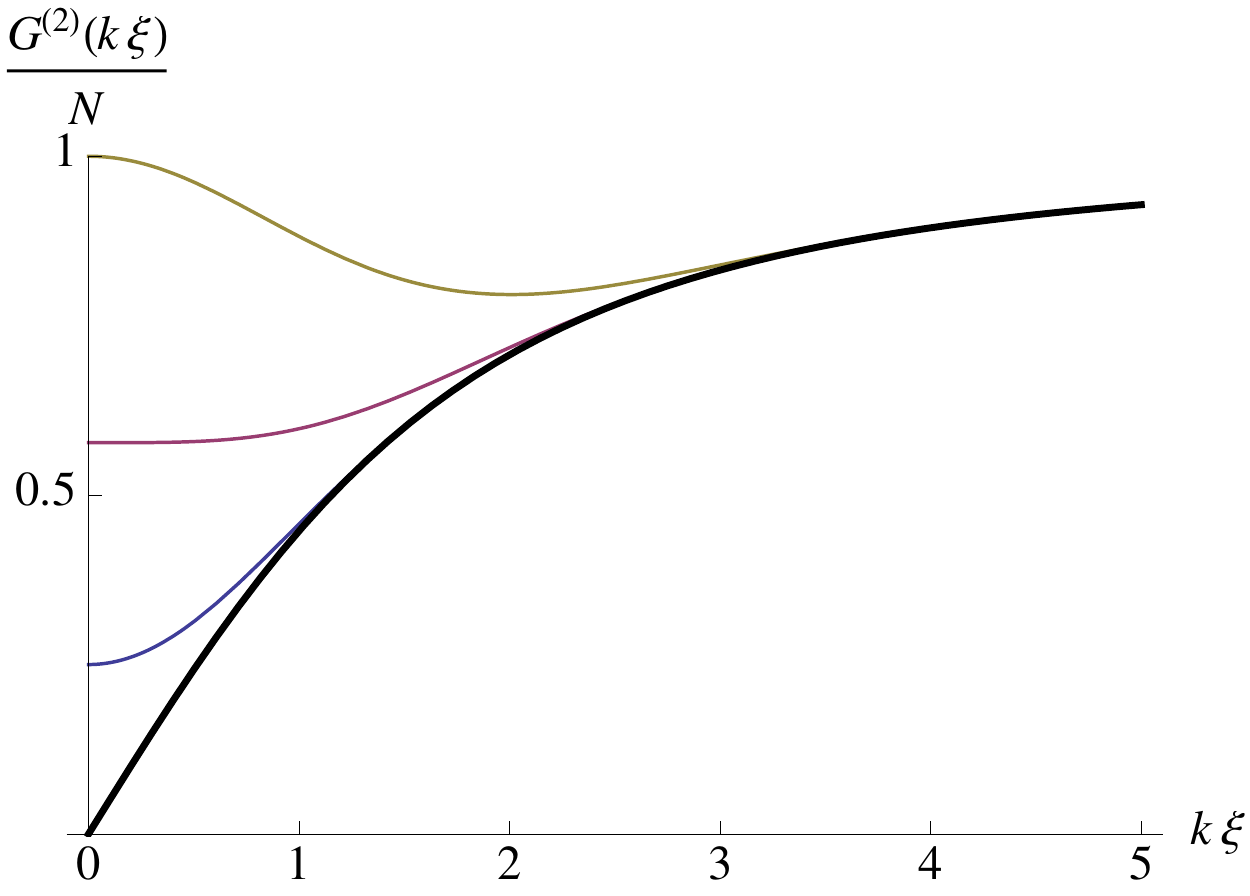} \, \includegraphics[width=0.45\columnwidth]{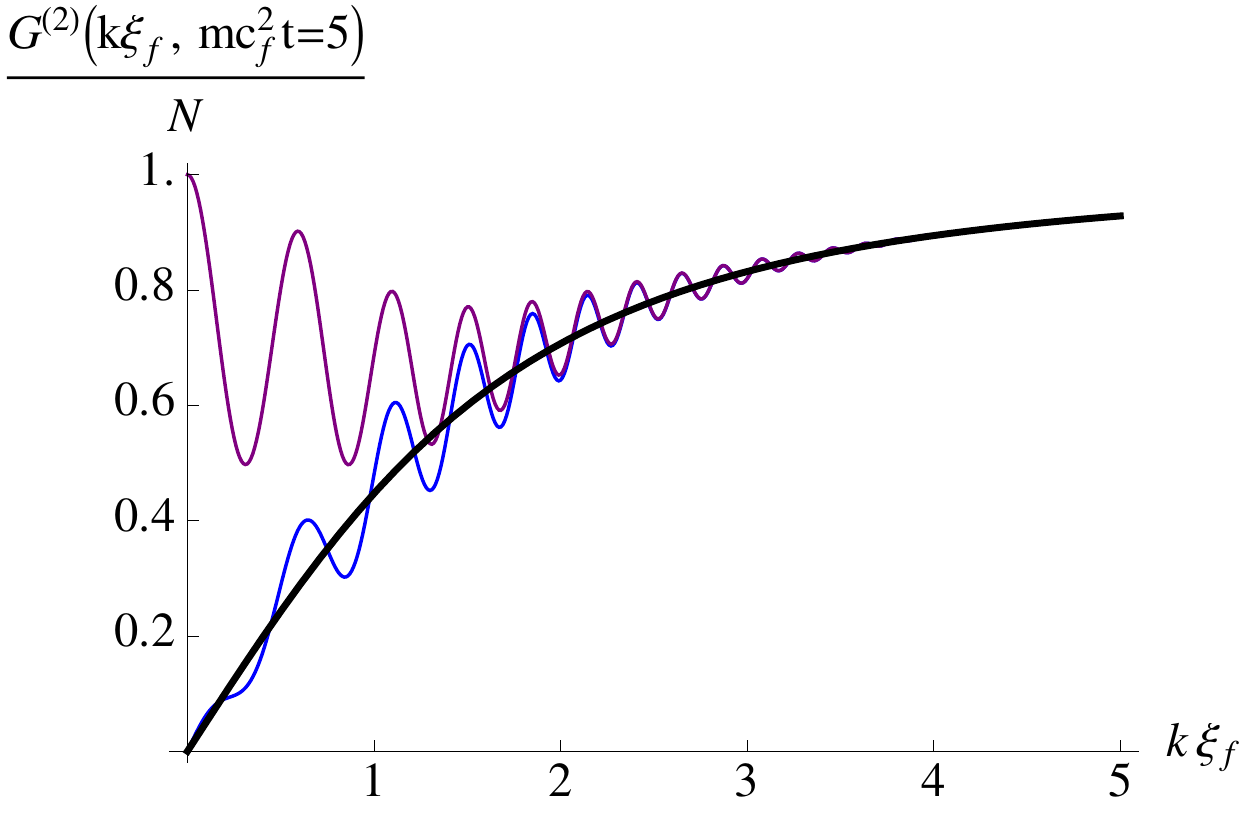}
\caption{Fourier transform of equal-time density-density correlation function.
On the left is shown $G^{(2)}(k)/N$ 
of \eq{density-density} 
when the phonon state itself is stationary and thermal, i.e. $c_k = 0$ and $2 n_k + 1 = \mathrm{coth}(\Omega_{k} 
/2T)$.
The various curves correspond to different temperatures: $T/mc^{2} = 0$ (black), $1/4$ (blue), $1/\sqrt{3}$ (purple) and $1$ (yellow).
On the right is shown $G^{(2)}(k)/N$ 
after a lapse of time $t$ following 
an increase in $c^{2}$ by a factor of $2$ (the same situation as in Figure 5 of \cite{Robertson-Michel-Parentani-2017}).
The two curves correspond to different initial temperatures: $T=0$ (blue) and $T=mc_{\rm in}^{2}$ (purple),
where $c_{\rm in}$ 
is the initial 
value of 
the phonon speed. 
As the lapse of time $t$ increases, 
at fixed $k$ $G^{(2)}(k)$ 
oscillates with frequency $2 \Omega_{k}$, 
and the number of oscillations visible in the plot of $G^{(2)}(k)/N$ 
increases.
\label{fig:G2k}}
\end{figure}

Measurements of $G^{(2)}(k,t)$ 
thus allow us to extract a certain amount of information about the phonon state.  
The first thing to notice is that, even in the absence of any phonons (i.e. $n_{\pm k} = c_{k} = 0$), $G^{(2)}(k)$ 
does not vanish but takes the value 
\begin{equation}
G^{(2)}_{\rm vac}(k) = N \left(u_{k} + v_{k} \right)^{2}\, , 
\label{G2-vacuum}
\end{equation}
 which is due to vacuum fluctuations, see the thick black curve in the left plot of Fig.~\ref{fig:G2k}. 
The presence of uncorrelated phonons (i.e. $n_{\pm k} \neq 0$, $c_{k} = 0$) increases this value by a relative amount of $n_{k} + n_{-k}$, see the colored 
curves in the left plot 
which correspond to 
thermal states with temperatures respectively equal to $T/mc^2 = 1/4$, $1/\sqrt{3}$ and $1$. 
In fact, in a thermal state, one has 
$n_{k}=n_{-k}=\left(e^{\Omega_{k} 
/T}-1\right)^{-1}$ where $T$ is the temperature
(in the condensate rest frame), 
and $G^{(2)}(k)$ thus becomes 
\begin{equation}
G^{(2)}(k) = N 
\left(u_{k}+v_{k}\right)^{2} \mathrm{coth}\left(\frac{\Omega_{k} 
}{2T}\right) \,. 
\label{G2-thermal}
\end{equation}
The ratio $G^{(2)}(k) / N$ is precisely the `static structure factor' shown 
(with $T/mc^2 = 0$ and $1/4$) 
in Fig. 7.4 of \cite{Pitaevskii-Stringari-BEC}
(for the sake of comparison their healing length is defined as $1/\left(\sqrt{2} m c\right)$, which is a factor of $1/\sqrt{2}$ smaller than ours given after Eq.~(\ref{disprel})). 
In the high-$k$ limit 
it tends to $1$,  
whereas in the low-$k$ limit it approaches $ 
T/mc^{2}$ 
(where we emphasise that, because of Galilean invariance, the $k \to 0$ limit gives the temperature in the rest frame of the condensate).  
By careful measurements of $\rho(x)$, it is thus possible to extract the physical quantities $N$, $\xi$ and $T$, as reported in~\cite{Schley-et-al}. 

It should be noticed that it is the total number of phonons $n_k + n_{-k}$ which is extracted, as we are unable to distinguish between 
the left- and right-moving 
sectors. Note that this is closely related to the fact that $\hat{\rho}_{k}$ and $\hat{\rho}_{k}^\dagger$ 
commute.  In effect, the indistinguishable character of $n_{k}$ and $n_{-k}$ is the price we pay 
in restricting ourselves to commuting measurements. 
Moreover, if we allow for anisotropic states where $n_{k}$ and $n_{-k}$ are characterized by different temperatures in the limit $k\to 0$, say $T_{\rm rm}$ and $T_{\rm lm}$, then the low-$k$ limit of $G^{(2)}(k)/N$ 
would yield the arithmetic mean of these two, i.e. it would approach
\begin{equation}
\frac{G^{(2)}(k)}{N} \to \frac{T_{\rm rm}+T_{\rm lm}}{2mc^{2}} \, .  
\label{G2rmlm}
\end{equation}

Finally, the presence of correlations (i.e. $c_{k} \neq 0$) causes the two-point 
function to vary sinusoidally with a frequency $2\Omega_{k}$, 
a relative amplitude 
$2 \left| c_{k} \right|$ and a phase equal to the phase of $c_{k}$, see the right plot of Fig.~\ref{fig:G2k} for two examples with the same values of $c_k$ but different occupation numbers. We refer to our former work~\cite{Robertson-Michel-Parentani-2017} where these two examples are obtained after having modified the trapping frequency 
in the perpendicular directions, $\omega_{\perp}$, in such a way that the square of the phonon speed $c^{2}$ increases by a factor of $2$, 
while starting with different initial temperatures. 
(Notice that in that work, we effectively used a different normalization convention for the Fourier transform of $\hat{\rho}$, so that there was no factor of $N$ out front in the expression for $G^{(2)}(k)$.)
The rate of change of $c^2(t)$ is 
chosen in such a way that it is slow with respect to $\omega_\perp$, 
so as 
not to cause any 
oscillations of the condensate itself; see Ref.~\cite{Robertson-Michel-Parentani-2017}. 
As explained in that work, 
longitudinal phonon modes with frequencies 
much lower 
than the rate of change of $\omega_{\perp}$ 
respond to that change as if it were 
sudden, 
inducing 
a significant mode amplification.
In quantum settings, when starting from vacuum, this dynamical Casimir effect (DCE) leads to the spontaneous 
production of maximally entangled pairs with opposite wave vectors $k,-k$, 
by which 
we mean that the maximal value of $|c_k|$ allowed by quantum mechanics (which characterizes 
pure states, 
see \eq{maxim}), is reached. 

\subsection{Assessing entanglement
\label{sub:HomoEntanglement}}  
We 
now turn to 
the extraction of 
the degree of entanglement of the state;  
the notions of nonseparability and steerability 
are briefly recalled in Appendix~\ref{app:Entanglement}.  

As mentioned in the introduction, 
inequality~(\ref{nonsep}) 
is a sufficient condition for nonseparability of the $k,-k$ bipartite state 
(and, in fact, is also necessary 
whenever the state is Gaussian) \cite{Campo-Parentani-2005,Adamek-Busch-Parentani,deNova-Sols-Zapata}. 
In Appendix~\ref{app:Entanglement}, we further 
demonstrate that a sufficient condition for (\ref{nonsep})
to be satisfied is  
\begin{equation}
G^{(2)}(k,t) < G^{(2)}_{\rm vac}(k) = N \left( u_{k} + v_{k} \right)^{2} \, , 
\label{strong_nonsep}
\end{equation}
for some time $t$.
This is one of our key results, and was previously 
reported (for isotropic states with $n_{k}=n_{-k}$) 
in \cite{Robertson-Michel-Parentani-2017}.
The indistinguishability of the $k$ and $-k$ sectors in the expression $\hat{\rho}_{k}(t) \hat{\rho}_{k}^{\dagger}(t)$ 
entering $G^{(2)}(k,t)$ 
causes the density-density measurements to act as an effective interferometer for these two channels.  With varying $t$, it gives us access to the two-mode phase space spanned by the modes $k$ and $-k$, see \eq{density-density}. 
Inequality~(\ref{strong_nonsep}), which expresses the (periodic) reduction of the 
noise below its vacuum value, implies the existence of a subfluctuant direction in this phase space~\cite{Campo-Parentani-2005}, thus directly revealing the entanglement of the state.  
This dipping below vacuum noise is a key feature of several practical measurements of entanglement, see also e.g. \cite{HOM,Lopes-et-al,Finazzi-Carusotto-2014}. 
Crucially, it is 
directly revealed by observations performed at a single 
time only; 
in particular, it does not yield any of the individual 
expectation values $n_{\pm k}$ or $c_{k}$, but only that they stand in a certain relation to one another.  
Indeed, the extraction of $n_{\pm k}$ and $c_{k}$ separately 
(which in the present settings are equivalent to the knowledge of the full covariance matrix, see Appendix~\ref{app:Entanglement}) 
{\it would} require the performance of non-commuting measurements, given that the number operators $\hat{n}_{\pm k} = \hat{b}^{\dagger}_{\pm k} \hat{b}_{\pm k}$ do not commute with $\hat{c}_{k} = \hat{b}_{k} \hat{b}_{-k}$, nor even the hermitian part of $\hat{c}_{k}$ with its anti-hermitian part.  

\begin{figure}
\includegraphics[width=0.45\columnwidth]{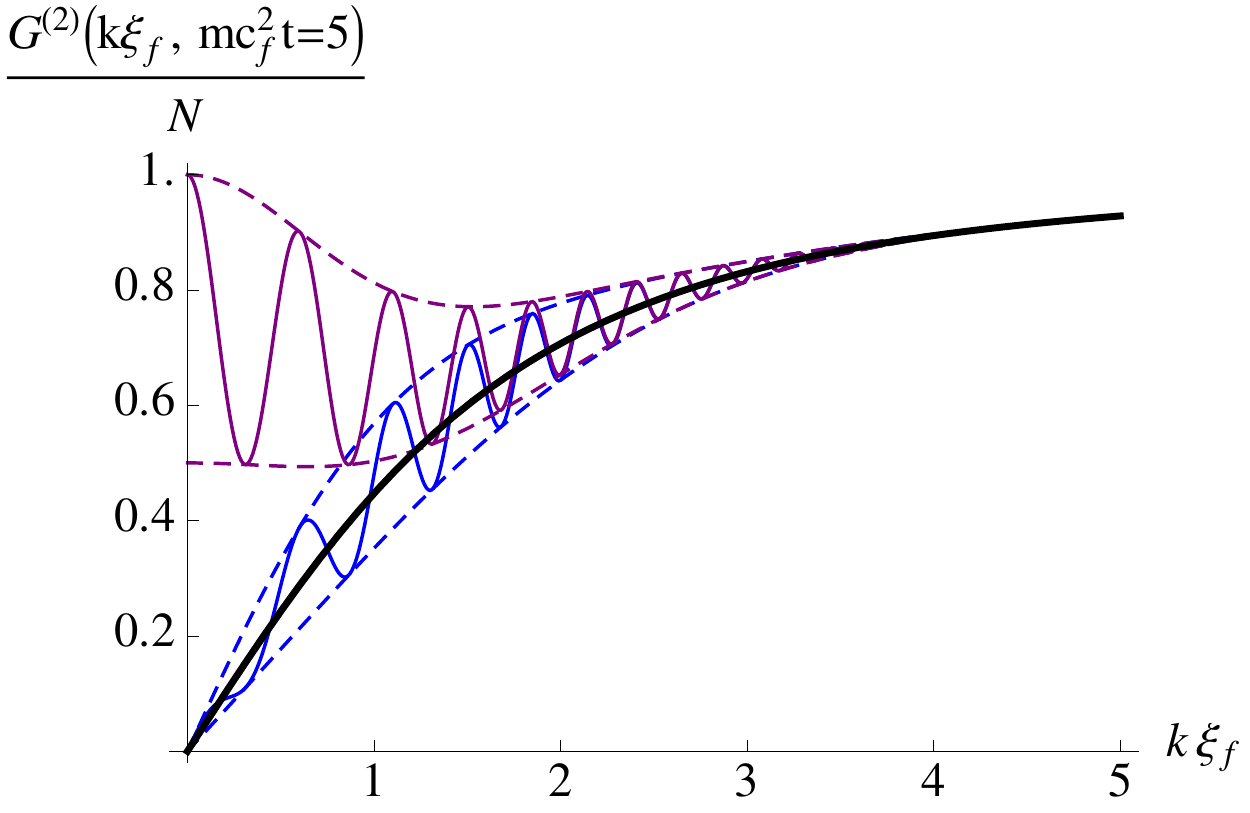} \, \includegraphics[width=0.45\columnwidth]{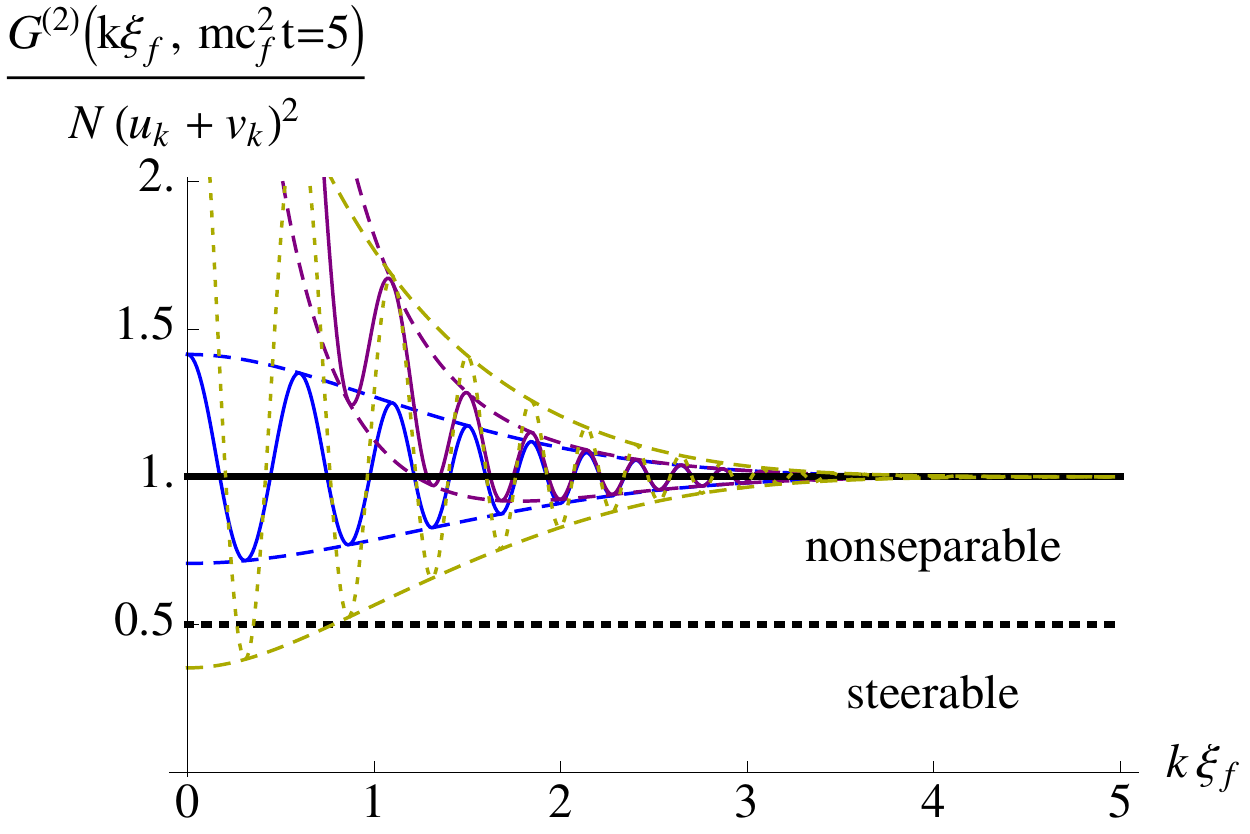}
\caption{Assessing the entanglement of the 
phonon state.
On the left are plotted the same density-density correlation functions as in the right panel of Fig.~\ref{fig:G2k}, with their upper and lower envelopes shown in dashed lines.  
Nonseparability is guaranteed 
when the minimum value of $G^{(2)}(k,t)$ 
is less 
than the correlation function associated to vacuum fluctuations, i.e. to the dipping of the lower envelope below the thick black curve. 
For the blue curve (at $T = 0$) the threshold is crossed 
for all $k$, whereas for the purple curve (at $T = mc_{i}^{2}$) it is crossed 
only for $k\xi \gtrsim 1.2$.
On the right are shown the same correlation functions, normalized by $\left(u_{k}+v_{k}\right)^{2}$ so that the nonseparability threshold occurs at exactly 
$1$.
Also included there in yellow is the result having increased 
$c^{2}$ by a factor of $8$ (at $T=0$).  For $k \xi \lesssim 0.8$, this curve satisfies the sufficient condition for steerability given in Eq.~(\ref{suff_steer}) 
and indicated by the thick horizontal dotted line. 
\label{fig:G2k_nonsep}}
\end{figure}

In practical terms, 
to verify if 
inequality~(\ref{strong_nonsep}) is satisfied, 
it suffices to look for the lower enveloping curve of $G^{(2)}(k,t)$, 
i.e., the 
minimum value reached by 
$G^{(2)}(k,t)$ 
when varying time. 
In the left plot of Fig.~\ref{fig:G2k_nonsep}, in dotted lines with the corresponding colors, 
we have added the upper and lower envelopes for the two examples considered in the right plot of Fig.~\ref{fig:G2k}. 
To facilitate the reading, 
on the right plot of Fig.~\ref{fig:G2k_nonsep} 
we represent the ratio $G^{(2)}(k,t)/G^{(2)}_{\rm vac}(k)$; 
the state is then 
nonseparable whenever the curve drops below $1$. 
One clearly sees that the lower envelope of 
the blue curve (which corresponds to the case with a vanishing  
temperature) 
is below the threshold for all values of $k$, as can be understood from the fact that 
the final phonon state 
in that case is a pure two-mode squeezed state. 
By contrast, 
the lower envelope of the purple curve (which corresponds to an initial temperature equal to $mc_{\rm in}^{2}$, where $c_{\rm in}$ is 
the initial phonon speed) dips 
below the nonseparability threshold 
only for $k\xi \gtrsim 1.2$. 

On the right plot of Fig.~\ref{fig:G2k_nonsep}, we have also added 
a 
thick dotted black line 
showing 
the 
threshold  
\begin{equation}
G^{(2)}(k,t) < \frac{G^{(2)}_{\rm vac}(k)}{2} \, , 
\label{suff_steer}
\end{equation}
which is a sufficient condition for steerability, 
see Appendix~\ref{app:Entanglement}. 
It is not crossed by 
either of 
the two cases shown on the left plot, 
although, 
for the case with a vanishing temperature (the blue curve), the phonon state is in fact 
steerable for all values of $k$. 
The reason is that there is a minimal value of $|c_k|$ below which the threshold~(\ref{suff_steer}) 
cannot be 
reached. 
To show that there is no problem of principle, we have added a third case (shown in yellow) 
which is obtained when varying the
trapping frequency $\omega_\perp$ in the perpendicular direction in such a way that the effective speed of sound $c^2$ appearing 
in \eq{disprel} changes by a factor of $8$, and not by a factor of $2$ as for 
the blue curve. 
This greater change induces enough mode amplification for certain modes (those with $k\xi \lesssim 0.8$) to dip below the 
threshold~(\ref{suff_steer})
, even though, as for the blue curve, the state is pure and all two-mode systems $(k,-k)$ 
are in fact steerable.  

\section{Inhomogeneous stationary background
\label{sec:Inhomogeneous}} 

\begin{figure}
\includegraphics[width=0.45\columnwidth]{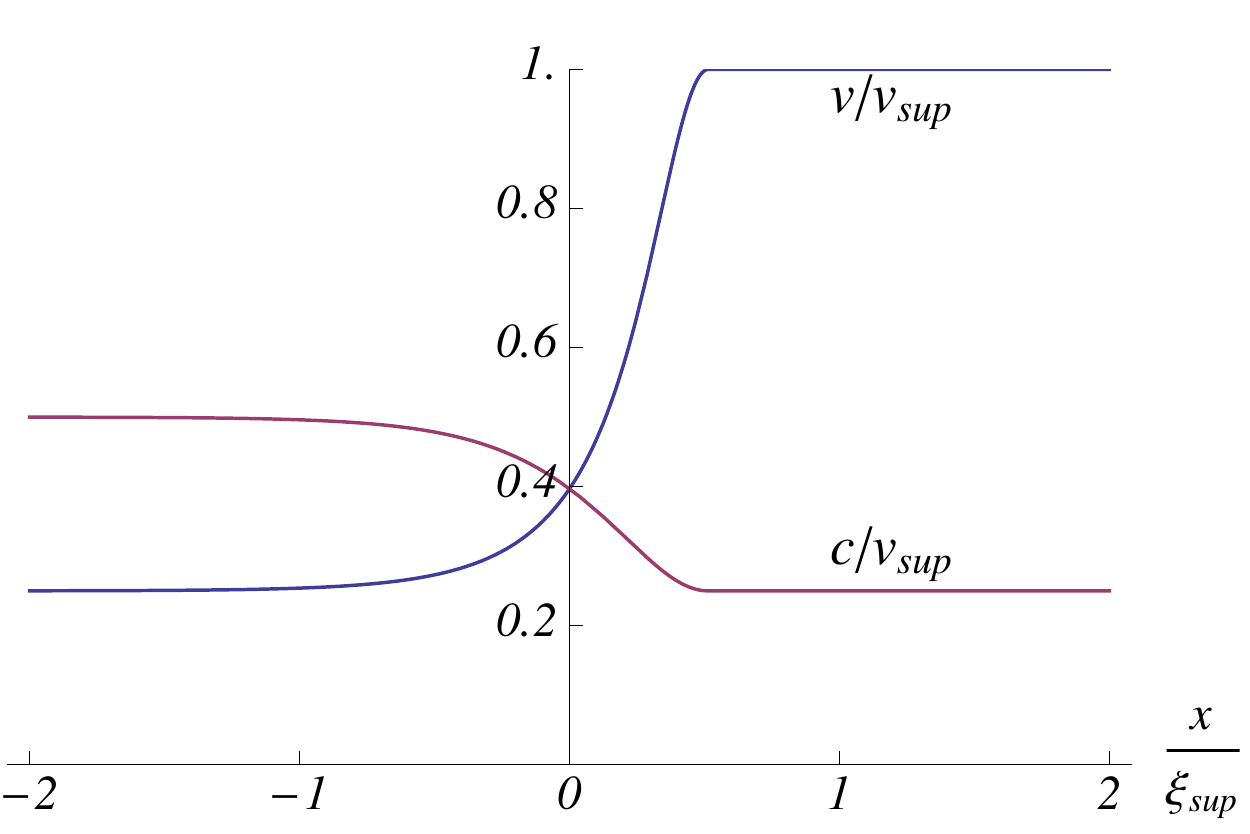} \, \includegraphics[width=0.45\columnwidth]{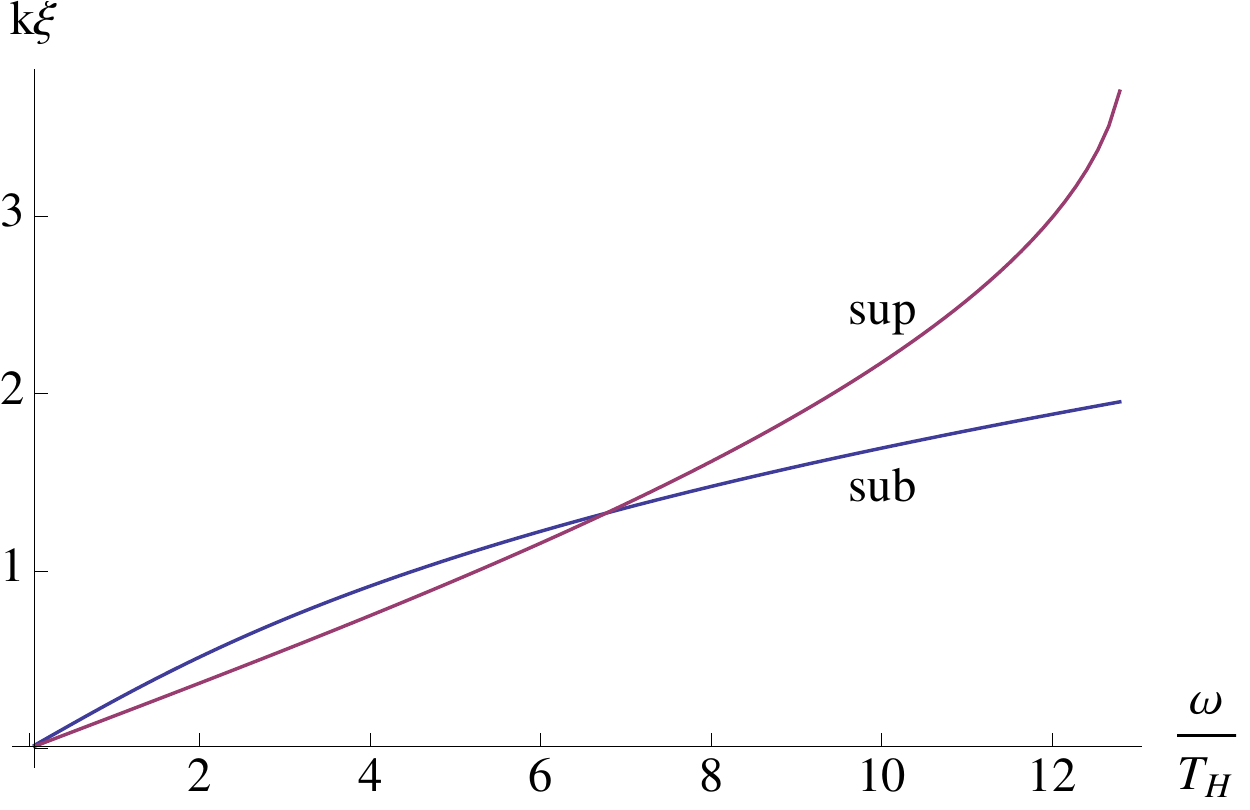}
\caption{Flow and dispersion profiles for a waterfall configuration. 
On the left are shown 
the flow velocity $v$ and the low-frequency phonon speed $c$, both normalized with respect to the supersonic (downstream) 
flow velocity $v_{\rm sup}$. 
In the supersonic 
region, the Mach number $M_{\rm sup}=4$ 
fixes that 
in the subsonic (upstream) region at 
$M_{\rm sub} = 1/2$, 
which is close to that reported in~\cite{Steinhauer-2016}; 
see also~\cite{Michel-Coupechoux-Parentani-2016}. 
The position is labelled such that $v$ and $c$ cross at $x=0$, which corresponds to the analogue event horizon.
On the right are shown the corresponding behaviors of the wave vectors of counter-propagating phonons (both in the subsonic and supersonic regions
and adimensionalized by the local value of the healing length) 
as functions of $\omega/T_{H}$, where $T_{H}$ is given in Eq.~(\ref{TH}).  
\label{fig:waterfall_v_c}}
\end{figure}

We now turn to the complementary case of  
a flow profile corresponding to an 
analogue black hole~\cite{Unruh-1981,AnalogueGravity-LivingReview}, 
by which 
we mean that 
the one-dimensional flow is transonic and stationary~\cite{Macher-Parentani-BEC}
and that the velocity increases in the direction of the flow. 
Hence the background is necessarily inhomogeneous 
but we shall also assume that it possesses (sufficiently long) homogeneous regions on both sides of the sonic horizon 
so that $\hat \rho_k$, the Fourier components of the density $\hat \rho(x)$, can 
be extracted on either side~\footnote{The assumption of homogeneity far 
from the horizon is rather mild 
as it has been shown that transonic flows analogous to black holes, i.e., 
with the flow velocity 
$v(x,t)$ increasing along the direction of the flow, obey ``no-hair'' theorems~\cite{Michel-Parentani-2015,Michel-Parentani-Zegers-2016}: 
they expel perturbations away from the sonic horizon where $v$ crosses the sound speed $c$. 
Therefore stationary asymptotically uniform transonic flows 
act 
(in a finite interval of $x$ containing the sonic horizon) 
as attractors for neighboring flows.}. 
An example of such a flow in atomic BEC 
is shown in the left panel of Fig.~\ref{fig:waterfall_v_c}, with the flow to the right so that $v>0$;  
it is close to that used in the experimental work~\cite{Steinhauer-2016} and to the examples shown in Figs.~1 and 2 of~\cite{Michel-Coupechoux-Parentani-2016}. 
It is now well-established that the transonic character of the background flow gives rise to a steady pair production of
outgoing phonons carrying opposite frequencies $\pm \omega$ 
(in similar fashion to 
the above described DCE 
which produces phonon pairs of opposite wave vectors $\pm k$). 
This steady emission 
is approximately thermal, with temperature (measured in the ``stationary'' frame in which the flow profile $v(x)$ is at rest) proportional to the rate of change of the total phonon speed at the sonic horizon (where $v = c$): 
\begin{equation}
T_{H} = \frac{1}{2\pi} \left. \partial_{x} \left(v-c\right) \right|_{\rm hor} \,, 
\label{TH}
\end{equation}
and can be understood in qualitative terms 
from the analogy with the Hawking radiation emitted by black holes~\cite{Unruh-1981}, 
or 
computed directly by solving the stationary Bogoliubov-de Gennes equation on such flows~\cite{Macher-Parentani-BEC}.
Once 
stationarity is achieved,  
the density-density 
correlation 
function~(\ref{G2}) 
depends only on the time difference $t-t'$, 
and not on $t$ and $t'$ individually. 
Given the form of $\hat{\rho}_{k}(t)$ in Eqs.~(\ref{densityFTphonon}) and~(\ref{phonon_operators}), 
this ensures that the expectation value of 
$\hat{\rho}_{k}(t) \hat{\rho}_{k'}^{\dagger}(t)$ 
in the asymptotic flat regions can be non-zero 
only when the corresponding frequencies $\omega_{k}$ and $\omega_{k'}$ 
(measured in the stationary frame) 
are equal in magnitude. 
As we shall see, the resulting correlation pattern projected onto 
the $(k,k')$-plane is 
more complicated than 
the 
strict 
$k=k'$ condition 
characterizing pair production 
in globally homogeneous flows.
\footnote{See~\cite{Boiron-et-al} for the structure of these correlations after time-of-flight measurements, 
and~\cite{Euve-et-al-2016} for the corresponding curves in a stationary inhomogeneous water wave system.} 

\subsection{Practical and conceptual difficulties}  

Let us briefly discuss the main properties characterizing density fluctuations in transonic flows, 
with particular emphasis on their differences with respect to the globablly homogeneous case, and the complications thereby induced.  

Firstly, 
at fixed $|\omega|$, three 
stationary modes are mixed by the scattering on a transonic flow~\cite{Macher-Parentani-BEC,Michel-Coupechoux-Parentani-2016},
rather than two as on a globally homogeneous background. 
One of these phonons is co-propagating with the flow, while the other two are counter-propagating; 
adopting a common notation, 
these shall be labeled 
by the superscripts $v$ and $u$, respectively. 
Taking the rest frame frequency $\Omega > 0$ (and given that the flow velocity $v>0$), the $v$-phonon has positive wave number $k$ while the $u$-phonons have negative $k$. 
More important are their (conserved) frequencies $\pm \omega$ in the stationary frame, which for $\Omega > 0$ 
give the signs of their energies: 
the $v$-phonon has positive energy while 
the two $u$-phonons have opposite energies. 
\footnote{Whereas the two positive-energy phonons can propagate throughout the entire space, 
the negative-energy phonon exists only 
in the region where the flow is supersonic; see Fig.~1 in Ref.~\cite{Busch-Parentani-2014} for the space-time trajectories followed by the 
three types of phonon.} 
Two types of phonon pair 
carrying zero total 
energy are 
thus 
spontaneously produced: $(u,u)$ pairs involving the two counter-propagating modes, 
and $(u,v)$ pairs involving the 
co-propagating $v$-mode.  

Despite 
this 3-mode mixing, it can be shown that the inequality
\begin{equation}
n^u_{\omega} \, n^u_{-\omega} - |c_\omega^{uu}|^2 < 0\, 
\label{nonsepinomega}
\end{equation}
guarantees that the bipartite state 
characterizing the two $u$-modes 
(and obtained 
by tracing over the 
$v$-mode) 
is nonseparable~\cite{Busch-Parentani-2014}. 
In strict analogy with the quantities entering Eq.~\eqref{nonsep}, we have 
$n^u_{\pm \omega} = \left\langle \hat{b}^{u\dagger}_{\pm \omega} \hat{b}^{u}_{\pm \omega} \right\rangle$ and $c_\omega^{uu} = \left\langle \hat{b}^{u}_{\omega} \hat{b}^{u}_{-\omega} \right\rangle$,
where $\hat{b}^{u}_{\pm \omega}$ $\left(\hat{b}^{u \dagger}_{\pm \omega}\right)$ destroys (creates) a $u$-phonon of frequency $\pm \omega$. 
It can also be shown that the couplings involving the $v$-mode are generally smaller than those relating the two $u$-modes~\cite{Busch-Parentani-2014}.
Therefore, 
the $v$-mode is essentially a spectator, and 
a fair understanding of the physics can be reached by focusing 
on the $(u,u)$ coupling terms. 
That said, from an experimental point of view, 
the indistinguishability of $n_{k}$ and $n_{-k}$ when measuring 
the $k$-th Fourier transform of 
density fluctuations 
means that 
one cannot simply discard the $v$-modes when one attempts 
to assess the nonseparability of the state, 
for the mean occupation numbers will be polluted by their presence. 

Secondly, and more crucially, is the fact that the 
$u$-phonons 
of opposite energy appear on opposite sides of the sonic horizon and then propagate 
away from each other in 
separate regions of space. 
As a result, one is now necessarily dealing with two distinct atomic 
densities: 
$\hat{\rho}^{\rm sub}(x)$ 
in the subsonic (upstream) 
region, and $\hat{\rho}^{\rm sup}(x')$ in the supersonic (downstream) region.
Explicitly, on Fourier transforming the density operator 
(and arbitrarily setting $t=0$ since the state is stationary and all measurements are made at equal time), 
we have 
(see \eq{densityFTphonon}
and~\cite{Steinhauer-2015}
) 
\begin{eqnarray}
\hat{\rho}_{k}^{\rm sub} &= &\sqrt{N^{\rm sub}} 
\left( u_{k}^{\rm sub}+v_{k}^{\rm sub} \right) \left( \hat{b}^{
{\rm sub}}_{k} + (\hat{b}_{-k}^{
{\rm sub}})^{\dagger} \right) \,,\nonumber\\
\hat{\rho}_{k'}^{\rm sup} &= &\sqrt{N^{\rm sup}} 
\left( u_{k'}^{\rm sup}+v_{k'}^{\rm sup} \right) \left( \hat{b}^{
{\rm sup}}_{k'} + (\hat{b}^{
{\rm sup}}_{-k'})^{\dagger} \right) \, .
\label{densityFTphonon-sub-sup}
\end{eqnarray}
We have added 
superscripts `sub' and `sup' 
to the Bogoliubov coefficients $u_{k}$ and $v_{k}$ to indicate that 
they depend 
on the healing lengths $\xi^{\rm sub}$ and $\xi^{\rm sup}$ 
defined 
on either side.  
We have also added 
superscripts `sub' and `sup' 
to the phonon operators $\hat b_k$ because they encode different modes
with support in non-overlapping regions of space; 
for instance, the commutator $[ \hat{b}^{{\rm sub}}_{k}, (\hat{b}^{
{\rm sup}}_{k'})^{\dagger}]$ 
vanishes even when $k=k'$. 
%
Finally, to facilitate the reading, we use an unprimed $k$ to refer to a measurement made in the subsonic region, and a primed $k'$ to refer to a corresponding measurement in the supersonic region.  It will prove convenient to employ this notation in the remainder of this section.  


The fact that the entangled 
outgoing 
$u$-phonons propagate in different regions has two important consequences.
On the one hand, they produce a non-local long-distance correlation pattern in the $(x,x')$-plane which signals the 
production of correlated phonon pairs of 
type $\left(u,u\right)$~\cite{Balbinot-et-al-2008,Carusotto-BH}. This non-local
correlation pattern has been reported in the experimental work~\cite{Steinhauer-2016} 
and agrees with theoretical predictions to a large extent~\cite{Michel-Coupechoux-Parentani-2016}.
On the other hand, as far as entanglement is concerned,
their associated density fluctuations no longer (locally) interfere.
This implies that one can no longer assess the entanglement 
of the state by simply comparing, 
as we did in \eq{strong_nonsep}, the measured value of $G^{(2)}(k)$ for some $k$ with the corresponding 
vacuum expression  $G^{(2)}_{\rm vac}(k)$. 

Instead, 
one is forced to follow the more indirect procedure proposed in~\cite{Steinhauer-2015}, 
which involves combining three different measurements of the density-density correlation function.  
Indeed, each of the three quantities entering Eq.~\eqref{nonsepinomega} should be estimated 
by pairing differently $\hat{\rho}^{\rm sub}_k$ and $\hat{\rho}^{\rm sup}_{k'}$ of the first and second
lines of \eq{densityFTphonon-sub-sup}, 
where $k$ and $k'$ are both negative (i.e. they correspond to $u$-modes) and are related by $\omega_{k}^{\rm sub} = -\omega_{k'}^{\rm sup}$;  
see Eq.~(\ref{disprel}) for the expression for $\omega_{k}$ (recalling that $c$ and $\xi$ are different in the two asymptotic regions), and the right panel of Figure~\ref{fig:waterfall_v_c} for the behavior of these solutions in the flow shown in the left panel. 
The wave numbers should thus be considered as functions of the frequency, which can be shown explicitly by writing $k=k^{u}_{\omega}$ and $k'=k^{\prime u}_{\omega}$.  
In short, 
the mean occupation numbers $n^u_{\omega} = n_{k^{u}_{\omega}}$ and 
$n^u_{-\omega} = n_{k^{\prime u}_{\omega}}$ 
should be extracted, respectively,
from the expectation values of $\hat{\rho}^{\rm sub}_{k^{u}_{\omega}} \hat{\rho}^{{\rm sub} \dagger}_{k^{u}_{\omega}}$ and $\hat{\rho}^{\rm sup}_{k^{\prime u}_{\omega}}\hat{\rho}^{{\rm sup} \dagger}_{k^{\prime u}_{\omega}}$, while the correlation term $c^{uu}_\omega$ should be extracted from the expectation value of 
$\hat{\rho}^{\rm sub}_{k^{u}_{\omega}} \hat{\rho}^{\rm sup}_{k^{\prime u}_{\omega}}$. 
\footnote{It thus appears that one needs to measure separately the three ingredients entering condition (\ref{nonsepinomega}). 
This raises a puzzling question, since on the one hand,
the occupation numbers $\hat n^u_{\pm \omega}= \hat{b}^{u\dagger}_{\pm \omega} \hat{b}^{u}_{\pm \omega}$, 
%
considered as operators, do {\it not} commute with the operator $\hat c_\omega^{uu} = \hat{b}^{u}_{\omega} \hat{b}^{u}_{-\omega}$, while on the other hand 
these three quantities are extracted from measurements (performed at a given time) of $\hat \rho(x)$ for various values of $x$ which do commute. The resolution of this paradox comes from the operator content of 
$\hat{\rho}^{\rm sub}_k$ and $\hat{\rho}^{\rm sup}_{k'}$ of \eq{densityFTphonon-sub-sup}.
Each of them contains {\it with equal weight} a destruction operator 
and a creation operator 
corresponding to modes 
with opposite 
values of 
$k$.
This guarantees that the three relevant combinations $\hat{\rho}^{\rm sub}_k\, \hat{\rho}^{\rm sub \dagger}_k$, 
$\hat{\rho}^{\rm sup}_{k'}\, \hat{\rho}^{\rm sup \dagger}_{k'}$
and $\hat{\rho}^{\rm sub}_k \hat{\rho}^{\rm sup}_{k'}$ 
commute with each other.
 Although they do not contribute to the three $uu$ quantities entering \eq{nonsepinomega}, 
the $v$-operators $(\hat{b}_{-k}^{v})^{\dagger}, (\hat{b}_{-k'}^{v})^{\dagger}$
are necessary to ensure the commutation of the density measurements.}

\subsection{Explicit expressions} 

We first consider 
the autocorrelation 
$G^{(2) {\rm sub}}(k) = \left\langle \hat{\rho}^{\rm sub}_{k} \hat{\rho}^{\rm sub \dagger}_{k} \right\rangle$ in the subsonic region, noting that the autocorrelation in the supersonic region is entirely analogous.  
Assuming the 
stationarity of the phonon 
state, we have
\begin{eqnarray}
G^{(2) {\rm sub}}(k,k) & = & N^{\rm sub} \left(u_{k}^{\rm sub} + v_{k}^{\rm sub}\right)^{2} \left( 1 + n_{k}^{\rm sub} + n_{-k}^{\rm sub} \right) \nonumber \\
& = & G^{(2) {\rm sub}}_{\rm vac}(k) \left( 1 + n_{k}^{\rm sub} + n_{-k}^{\rm sub} \right) \,.
\label{density-density2}
\end{eqnarray}
As in \eqref{strong_nonsep}, we see 
the key role 
played by the vacuum two-point function $G^{(2) {\rm sub}}_{\rm vac}(k)$ 
in extracting the observable quantity, here $n^{\rm sub}_{k} + n^{\rm sub}_{-k}$, characterizing the phonon state. 
As could have been expected from 
the analysis of the former section,
it is the total occupation number $n_{k}^{\rm sub}+n_{-k}^{\rm sub}$ that is extracted from the density 
measurements. 
Hence the $v$-modes do (positively) contribute to the measurements of $\hat{\rho}^{\rm sub}_k \, \hat{\rho}^{\rm sub \dagger}_k$ 
which therefore
only gives an upper 
bound 
for 
the $u$-mode occupation number. 
Note also that, if the state is stationary,
then since $\left|\omega^{u}_{k}\right| \neq \left|\omega^{v}_{k}\right|$ whenever the flow velocity is non-zero, 
we necessarily have on each side $c_{k}=0$, 
i.e. $u$- and $v$-modes of wave numbers $k$ and $-k$ are completely uncorrelated.  
%
Whereas, in the homogeneous case, the interference between phonon modes of opposite wave numbers
is useful because it combines precisely those two modes that are entangled by the time-varying background, 
here a possible 
interference (for instance due to some lack of stationarity of the background flow) 
would be a hindrance in that it combines modes which are not related by the 
stationary analogue 
Hawking effect, and thus 
would 
pollute the measurements.  
\footnote{For the sake of clarity, we wish to emphasize that the $v$-modes which are coupled to the $u$-modes by the density measurements, and which are related to them through having the same magnitude of wave number $|k|$, are to be distinguished from the $v$-modes which are coupled to the $u$-modes by the analogue Hawking effect, related by having the same magnitude of frequency $|\omega|$, and which only appear in the downstream (supersonic) region; see Eq.~(58) of~\cite{Macher-Parentani-BEC}.  In the homogeneous case, by contrast, the modes coupled by the DCE are {\it precisely} those which are also coupled by the density measurements.}  

It turns out that extracting the correlation term $c^{uu}_{\omega}$ is somewhat simpler, when 
the stationarity of the 
phonon 
state is assumed 
and when the sub- and supersonic regions are sufficiently well-separated that the residual contribution of the other modes (including $v$-modes) can be safely ignored. 
Letting $k = k^{u}_{\omega}$ and $k'=k^{\prime u}_{\omega}$, 
one easily verifies that only one term 
in the cross-correlation 
is non-vanishing: 
\begin{equation}
G^{(2) {\rm sub/sup}}(k,-k') = 
\left\langle \hat{\rho}^{\rm sub}_{k} \hat{\rho}^{\rm sup}_{k^{\prime}} \right\rangle = 
\sqrt{N^{\rm sub} N^{\rm sup}} 
 \left(u_{k}^{\rm sub} + v_{k}^{\rm sub}\right) \left(u_{k^{\prime}}^{\rm sup} + v_{k^{\prime}}^{\rm sup}\right) c_{k,k^{\prime}} \,, 
\end{equation}
where we have defined $c_{k,k^{\prime}} \equiv \left\langle \hat{b}_{k}^{u, {\rm sub}} \hat{b}_{k^{\prime}}^{u, {\rm sup}} \right\rangle$.  

In brief, under the assumption of stationarity, noticing that the populations $n^{u}_{\pm \omega}$ 
are both being overestimated 
by the $G^{(2)}$ measurements 
if one ignores the population of the $v$-modes, 
\eq{nonsepinomega} combined with the above equations tells us that 
a sufficient criterion for 
nonseparability is
\begin{equation}
\Delta^{(2)}(k,k^{\prime}) \equiv [G^{(2) {\rm sub}}(k,k) - G^{(2) {\rm sub}}_{\rm vac}(k)] \, [G^{(2) {\rm sup}}(k',k') - G^{(2) {\rm sup}}_{\rm vac}(k')]- |G^{(2) {\rm sub/sup}}(k,-k')|^2 < 0 \,  . 
\label{G2_difference}
\end{equation}

\begin{figure}
\includegraphics[width=0.45\columnwidth]{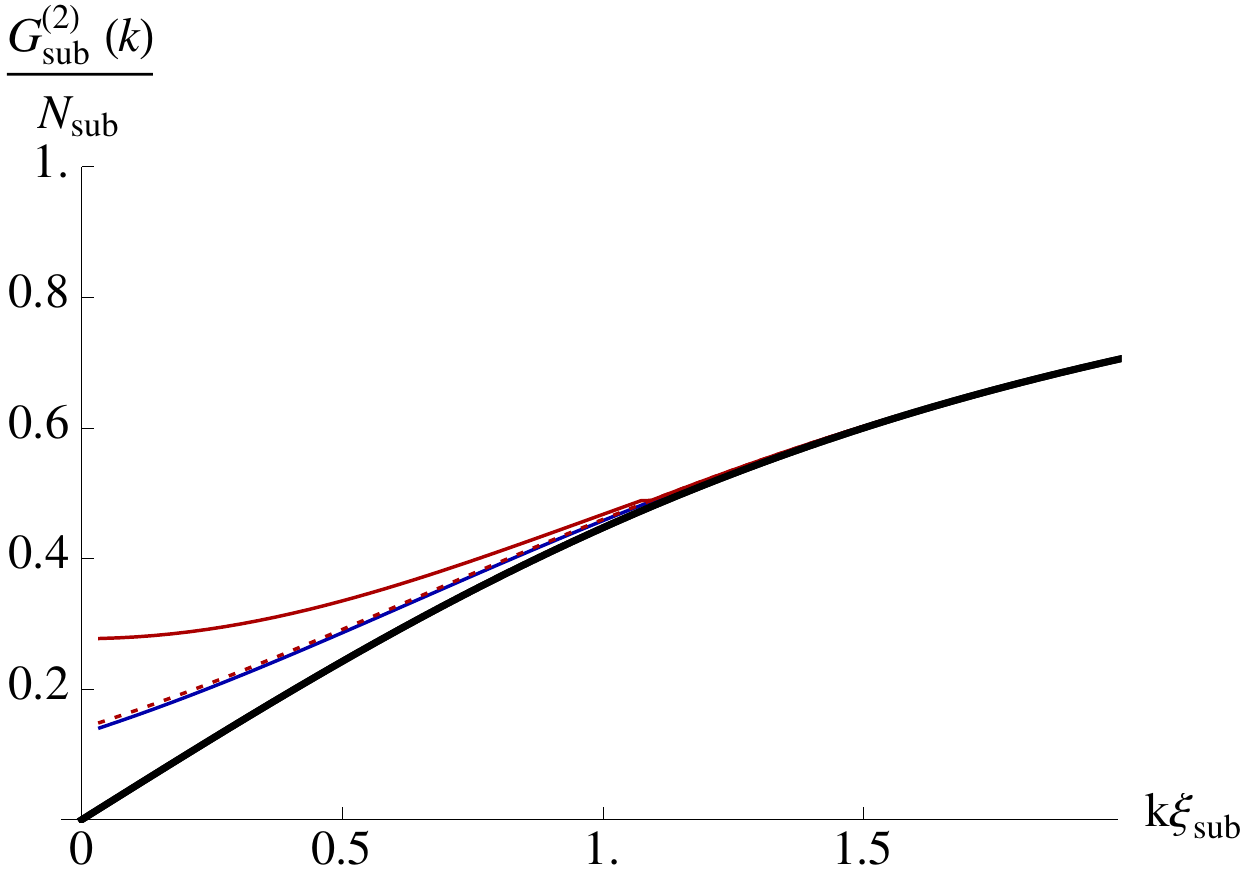} \, \includegraphics[width=0.45\columnwidth]{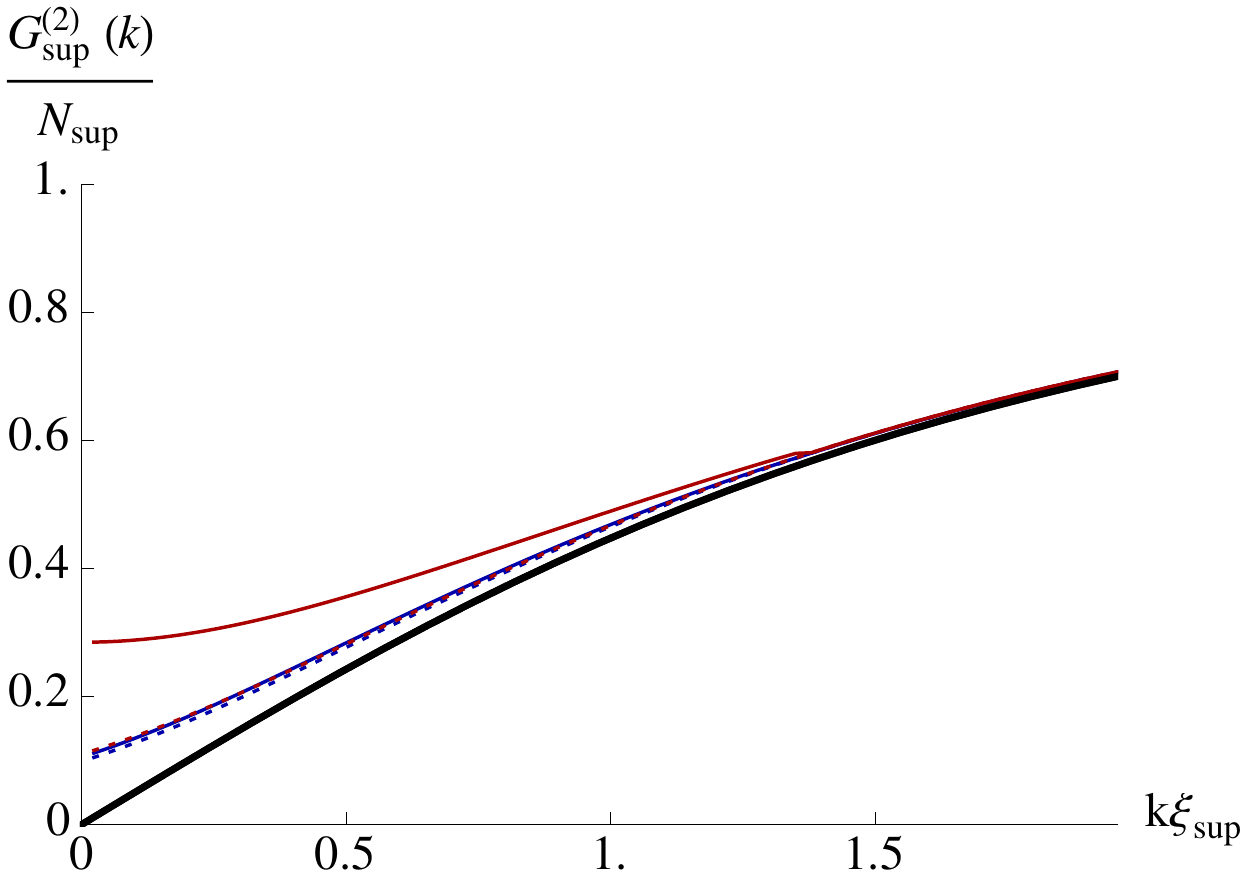} \\
\includegraphics[width=0.45\columnwidth]{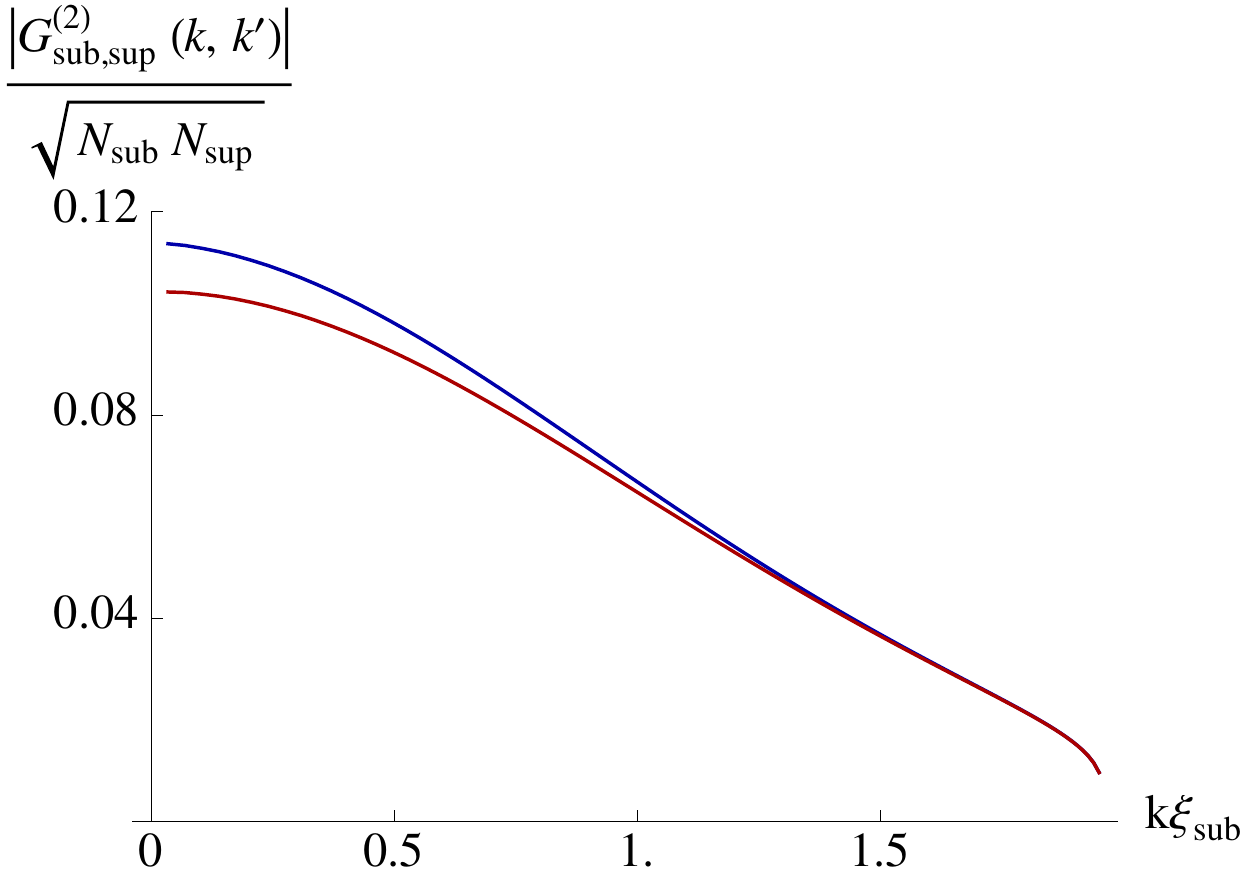} \, \includegraphics[width=0.45\columnwidth]{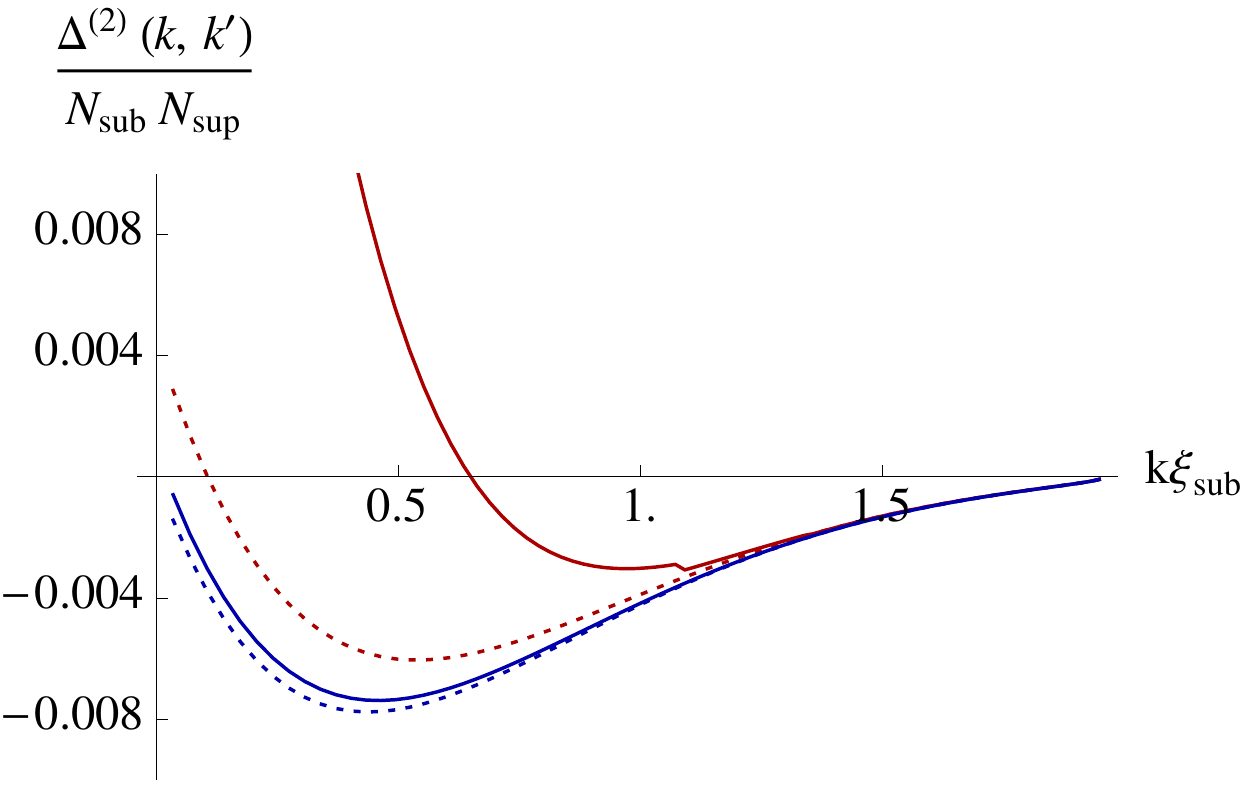}
\caption{Density-density correlation function for a waterfall in BEC, with Mach number in the supersonic region $M_{\rm sup}=4$ (close to that reported 
in~\cite{Steinhauer-2016}). 
The plots in the upper row show the autocorrelation $G^{(2)}(k)$ in the subsonic (left panel) and supersonic (right panel) regions, 
while the lower left panel plots the cross-correlation $G^{(2)}(k,k')$. 
The lower right panel shows instead 
the difference appearing in inequality~(\ref{G2_difference}).  
The variously colored curves correspond to different initial temperatures (in the frame of the condensate) of the incident $v$-modes: $T_{\rm in} = 0$ (blue curves) and $T_{\rm in} 
= 2 T_{H}$ (red curves -- the kinks appearing in these are numerical artifacts due to assuming that the scattering is trivial above $\omega_{\rm max}$, the maximum frequency at which $(u,u)$ phonon pairs are produced).  
In the upper plots, the thick black curves show the autocorrelation in vacuum, which must be subtracted before insertion into inequality~(\ref{G2_difference}).  
The solid curves are those that are actually observed; by contrast the 
dotted curves are those that would result if the measurements were able to distinguish $u$- and $v$-modes, and only the occupation numbers of $u$-modes were to appear in all expressions.  
When the initial temperature $T_{\rm in} = 2 T_{H}$, the pollution of the density measurements by the presence of $v$-modes dramatically affects the observations of $G^{(2)}(k)$ in the subsonic and supersonic regions (and, therefore, also the computed value of $\Delta^{(2)}$), as can be seen by comparing the solid and dotted red curves in the upper plots and in the lower right plot.  
Indeed, since the $u$-$v$ coupling is small, the occupation numbers of $u$-modes remain essentially the same whether or not there is an initial thermal state, as indicated by the fact that the dotted red curves are almost the same as the blue curves in the upper plots. 
\label{fig:G2_Tzero}}
\end{figure}

To illustrate what this procedure entails, 
we have plotted in Fig.~\ref{fig:G2_Tzero}
the three relevant Fourier transforms of $G^{(2)}(x,x')$,
as well as the difference of~(\ref{G2_difference}), 
that are theoretically obtained by solving the BdG equation
on a stationary transonic flow described by an exact solution (called a ``waterfall'' or ``half-soliton'') 
of the GPE on a step function potential~\cite{Michel-Coupechoux-Parentani-2016}. 
These stationary asymptotically homogeneous flows form a one-parameter family 
of solutions that can be labeled by the asymptotic Mach number $M_{\rm sup} = v_{\rm sup}/c_{\rm sup}$ 
in the downstream supersonic region; the corresponding Mach number in the upstream subsonic region is $M_{\rm sub} = M_{\rm sup}^{-1/2}$. 
We worked with $M_{\rm sup} = 4$ 
as it matches what has been observed in~\cite{Steinhauer-2016}. We also considered two initial temperatures for the incident $v$-modes: 
$T_{\rm in}^{\Omega} = 0$ and 
$T_{\rm in}^{\Omega} 
= 2 T_{H}$,
where $T_{H}$ (of Eq.~(\ref{TH})) 
is the effective low-frequency temperature of the emitted 
$u$-phonons.

In terms of the $u$- and $v$-modes that combine to give the autocorrelation of Eq.~(\ref{density-density2}), Hawking radiation is inherently anisotropic: the $u$- and $v$-mode populations on any one side are generally very different.  
If the ingoing $v$-mode state is close to vacuum, the $k \to 0$ limit of $G^{(2)}(k)$ (governed by Eq.~(\ref{G2rmlm})) will then be $T^{u}_{\Omega}/2mc^{2}$. 
Notice that 
the extracted temperature of the $u$-modes 
is that measured in the condensate rest frame, i.e. it is associated to the rest frame frequency $\Omega$ rather than to the conserved frequency $\omega$, these being related by a Doppler shift, see Eq.~(\ref{disprel}).~\footnote{In fact, the inverse transformation should be considered to express the initial temperature of the condensate $T_{\rm in}$, naturally expressed in terms of $\Omega$, in terms of the conserved frequency $\omega$ governing the Hawking effect; see Sec.~II C of~\cite{Busch-Parentani-2014}.}  
In the limit $k\to 0$, they become proportional, namely $\Omega = \omega/\left(1-M\right)$ where $M=v/c$ is the Mach number.  Therefore, 
the associated temperatures are related in the same manner:  
we have, for the counter-propagating $u$-modes,
\begin{equation}
T^{u}_{\Omega} = \frac{T^{u}_{\omega}}{\left| 1-M \right|} \,.
\label{tempshift}
\end{equation}
Taking into account the above facts that, when 
the ingoing state is vacuum, in the subsonic region 
the $v$-modes are in their vacuum state and the outgoing $u$-modes have an $\omega$-temperature close to $T_{H}$ of Eq.~(\ref{TH}), 
we arrive at the conclusion that the $k\to 0$ limit of $G^{(2)}(k)$ is 
\begin{equation}
\frac{G^{(2)}_{\rm sub}(k)}{N_{\rm sub}} \to \frac{1}{2} \, \frac{T_{H}}{mc_{\rm sub}^{2}} \, \frac{1}{1-M_{\rm sub}} \,.  
\label{G2limit}
\end{equation}
As far as we know, this combination of effects has not yet been discussed in the literature, 
although it can be shown to agree with Eq.~(35) of~\cite{Michel-Coupechoux-Parentani-2016}. 
We should also point out that, for the waterfall flow (close to the flow realised in~\cite{Steinhauer-2016}) with $M_{\rm sub} = 1/2$, 
the product $2 \left(1-M_{\rm sub}\right) = 1$.  In that case, we recover from Eq.~(\ref{G2limit}) the standard 
expression $G^{(2)}(k)/N \to T_{H}/mc^{2}$ (see below Eq.~(\ref{G2-thermal})), 
as if neither of the above effects were modifying $G^{(2)}(k)$. 
Notice finally that a more complicated expression than~(\ref{G2limit}) governs the low-$k$ behavior of $G^{(2)}(k)$ in the supersonic region due to the production of $v$-modes; see the first term in Eq.~(53) of~\cite{Macher-Parentani-BEC}. 

\section{Conclusion
\label{sec:Conclusion}}

In this paper, 
we have considered the properties of the equal-time density-density correlation function in homogeneous domains of elongated effectively one-dimensional atomic cold gases. 
We began by studying globally homogeneous systems, 
examining 
the time-dependence of the 
Fourier transform of the density-density correlation function 
at fixed wave number $k$. 
In this case, 
$G^{(2)}(k,t)$, the $k$-th component of the two-point function, displays periodic oscillations whose amplitude depends on the strength of the correlations between the phonon modes of opposite wave numbers. These oscillations arise from the interference 
between the contributions of these 
modes when measuring the atomic density. We demonstrated that if the minimal value periodically reached by $G^{(2)}(k,t)$ goes below $G^{(2)}_{\rm vac}(k)$, the constant value $G^{(2)}(k)$ has in the vacuum state, 
then the bipartite phonon state of wave numbers $k,-k$ is necessarily nonseparable. Remarkably, repeated {\it in situ} measurements of the atomic 
density $\rho(x)$ at a given time (one for which $G^{(2)}(k,t) < G^{(2)}_{\rm vac}(k)$) 
are sufficient to assess the nonseparability of the bipartite phonon state. In this we see that the commuting character of these measurements does not prevent one from having access to the entanglement of the state, in contrast to 
the fact that the knowledge of all entries of the covariance matrix does require the performance of 
non-commuting measurements. In fact, a closer analysis shows that {\it in situ} measurements of $\rho(x)$ performed at an 
appropriate time give us access to the particular combination of the elements of the covariance matrix which governs the degree of entanglement among the two parties. In mathematical terms, the combination appearing in \eqref{nonsep} gives the lowest eigenvalue of the determinant used in the function ${\cal P}$ which appears 
in the generalized Peres-Horodecki criterion. Hence the sign of the difference 
entering \eqref{nonsep} is equal to 
the sign of ${\cal P}$ (see Appendix~\ref{app:Entanglement}). 
By a similar analysis (performed at the end of Sec.~\ref{sub:HomoEntanglement}), we showed that the stronger criterion of steerability can also be experimentally verified by studying the behavior of $G^{(2)}(k)$ at a given time, i.e. by commuting measurements of density fluctuations. 
We hope that forthcoming experiments could exploit the simple analysis of Sec.~\ref{sub:HomoEntanglement} to assess entanglement of bipartite phonon states $(k,-k)$. 

In the second part of the paper, we studied the density fluctuations measured asymptotically on each side of 
a transonic stationary flow whose 
velocity increases along the direction of the flow, mimicking 
a black hole metric and giving 
rise to the steady production of phonon pairs of zero total energy 
by a process analogous to the Hawking effect. The relevant phonons, i.e. those which are counter-propagating with respect to the flow, are emitted on either side of the horizon, so that (in stark contrast to the homogeneous case) they live in non-overlapping regions of space and do not (directly) interfere. 
It is thus 
no longer possible to extract information about the entanglement of the relevant phonon modes 
by directly observing the oscillations of the density-density correlations. Rather, one should combine three different measurements of these correlations. Two of them are autocorrelations performed in each asymptotic region, 
and give upper bounds for the mean occupations numbers $n^u_{\pm \omega}$ of the relevant phonons. The reason they give only 
upper bounds is our inability to subtract the 
(weak but unknown) contributions of the co-propagating modes, which act as spectator modes in the present process in that they are only weakly involved in the mode mixing taking place near the sonic horizon. The third measurement instead is a cross-correlation, coming 
from Fourier components of the two-point function evaluated on either side of the horizon. It gives a measure of the norm of $c_\omega^{uu}$ which governs the strength of the correlations between the two relevant phonon modes. 
Although the spectator co-propagating modes do not contribute to the mean value of this measurement (because we assumed stationarity), 
their contribution to the operators 
being measured guarantees 
that the third measurement commutes with the former two.
To us this is a remarkable illustration of the counter-intuitive nature 
of quantum mechanics. 


\section*{Acknowledgments}

We are grateful to Jeff Steinhauer, Ted Jacobson, Bill Unruh, Chris Westbrook and Carsten Klempt 
for interesting discussions that provided motivation for the writing of this paper.  
We also thank the organizers of the 21st Peyresq Physics meeting in June 2016 where some of the discussions took place. 
This work was supported by the French National Research Agency through 
the Grant No. ANR-15-CE30-0017-04 associated with the project HARALAB,
and by the Silicon Valley Community Foundation 
through their grant FQXi-MGB-1630 ``Observing the entanglement of phonons in atomic BEC''. 


\begin{appendices}
\numberwithin{equation}{section}  

\section{Degrees of entanglement in bipartite systems
\label{app:Entanglement}}  

Entanglement is one of the most telling signs of the quantum nature of physics. 
Moreover 
it is a rich and often subtle subject. 
Whereas for pure states `entanglement' is a fairly well-defined property, with several equivalent formulations, it was pointed out by Werner \cite{Werner} 
and subsequent authors that for mixed states these formulations are no longer equivalent.  This leads to a hierarchy of different degrees of entanglement for general quantum states.  Here 
we consider two such notions: {\it nonseparability} and {\it steerability}.
In preparation for the study of density perturbations in atomic Bose gases presented in the main text, we 
restrict our attention to two-mode bosonic states, each single-mode subsystem having its own quantum amplitude operators $\hat{b}_{j}$ and $\hat{b}_{j}^{\dagger}$ ($j = 1,2$) subject to the usual bosonic commutation relation $\left[ \hat{b}_{i} \,,\, \hat{b}_{j}^{\dagger} \right] = \delta_{i,j}$.  

\subsection{
The criteria}

\subsubsection{Nonseparability}

Nonseparable states are best defined as the complement of the set of 
separable states, for which an explicit 
definition can be given. The 
bipartite 
state $\hat{\rho}_{1,2}$ is said to be {\it separable} whenever it can be written in the form
\begin{equation}
\hat{\rho}_{1,2} = \sum_{a} P_{a} \, \hat{\rho}_{1}^{a} \otimes \hat{\rho}_{2}^{a} \,,
\label{separable}
\end{equation}
where $\hat{\rho}_{j}^{a}$ are single-mode states and the $P_{a} \geq 0$ are real numbers.  
Then $\sum_{a} P_{a} = 1$, and the state $\hat{\rho}_{1,2}$ 
has 
the properties of a probability distribution. When more than one $P_a \neq 0$,  $\hat{\rho}_{1,2}$
entails 
correlations between the 
subsystems $1$ and $2$, 
but only in a classical sense: 
the overall state can be obtained 
by using a random number generator to pick the 
factorized 
$a^{\rm th}$-state 
$\hat{\rho}_{1}^{a} \otimes\hat{\rho}_{2}^{a}$ 
distributed according to the probability distribution $P_{a}$, and 
placing each subsystem 
separately in the states $\hat{\rho}_{1}^{a}$ and $\hat{\rho}_{2}^{a}$.  Note that the set of all separable states contains the set of all uncorrelated states as a subset.  Conversely, 
{\it nonseparable} states are those which cannot be written in the form (\ref{separable}).  Such states are necessarily correlated, but their correlations cannot be accounted for by 
the above 
classical means.


\subsubsection{Steerability}

The notion of steerability was originally formulated along the lines of thinking present in the original EPR paper~\cite{EPR}.
The idea is to consider making measurements on one subsystem, say $1$, and using the results of such measurements to infer the values of correlated quantities for the second subsystem $2$. 
\footnote{Schr\"{o}dinger~\cite{Schroedinger-1935} coined the term {\it steering} to describe 
the ability to `steer' the 
subsystem $2$ into an eigenstate of either of two non-commuting observables $\hat A_2,\hat B_2$, 
by choosing 
the measurement made on the subsystem $1$.} 
As formulated more recently by Reid~\cite{Reid-1989}, the mathematical description of steering can be written as
\begin{equation}
\Delta_{\rm inf} A_{2} \cdot \Delta_{\rm inf} B_{2} < \frac{1}{2} \left| \left\langle \left[ \hat{A}_{2}, \, \hat{B}_{2} \right] \right\rangle \right|_{\rm min} 
\,,
\label{steering_defn}
\end{equation}
where $\Delta_{\rm inf} A_{2}$ refers to the inferred standard deviation of $A_{2}$ on subsystem 2 having made a corresponding measurement on subsystem 1, that is,  
\begin{equation}
\Delta_{\rm inf} A_{2} = \sqrt{ \left\langle (\hat A_{2} - \bar{A}_{2}(A_1) 
)^{2}  \right\rangle} \, ,
\label{inf-dev}
\end{equation}
where $\bar{A}_{2}(A_{1})$ 
is the conditional (mean) value of $\hat A_{2}$ given that 
a measurement of $\hat A_{1}$ on subsystem 1 yields the eigenvalue $A_1$.  
Notice that a measurement of $\hat B_{1}$, which does not commute with $\hat{A}_{1}$, 
is performed on subsystem 1 when computing $\Delta_{\rm inf} B_{2}$. 
The right-hand side of~(\ref{steering_defn}) 
is the minimum value of the product of standard deviations according to the Heisenberg uncertainty principle applied to system $2$. 
The state is steerable, then, when correlations between the two subsystems are so strong that the inferred standard deviations are able to violate the usual uncertainty relation.  Note that, generally speaking, the steerability criterion is asymmetric: it is different depending on which of the two subsystems is measured and which is being `steered'.
Notice also that 
steerability is a stronger entanglement criterion than nonseparability; 
see the forthcoming analysis and Refs.~\cite{Wiseman-Jones-Doherty-PRL,Jones-Wiseman-Doherty-PRA}.  

\subsection{Sufficient inequalities} 

The mathematical definitions of nonseparability and steerability given above are too general 
to be amenable to be compared with experimental data. 
It is then useful to identify inequalities relating observable quantities which, when violated, 
are sufficient to assess that the state be entangled according to one of the above criteria.   

\subsubsection{Nonseparability}

%
 
A commonly used sufficient criterion for nonseparability is the generalized Peres-Horodecki (gPH) criterion, an algebraic condition on the covariance matrix of a two-state system.  It is expressed by Simon~\cite{Simon} using 
the two-mode operators $\hat{X} = \left[ \hat{q}_{1},\hat{p}_{1},\hat{q}_{2},\hat{p}_{2} \right]$ and the covariance matrix $V_{\alpha\beta} = \left\langle \left\{ \Delta \hat{X}_{\alpha}, \Delta \hat{X}_{\beta} \right\} \right\rangle /2$, 
where, as usual, $\hat{q}_{j} = \left(\hat{b}_{j} + \hat{b}_{j}^{\dagger}\right)/\sqrt{2}$ and $\hat{p}_{j} = \left(\hat{b}_{j} - \hat{b}_{j}^{\dagger}\right)/\left(\sqrt{2}\, i\right)$ 
for each subsystem. 
A necessary condition for separability is
\begin{equation}
V + \frac{i}{2} \left( \begin{array}{cc} J & 0 \\ 0 & -J \end{array} \right) \geq 0 \,,\qquad {\rm where} \qquad J = \left( \begin{array}{cc} 0 & 1 \\ -1 & 0 \end{array} \right) \,.
\label{gPH}
\end{equation}
That is, if the state is separable, the operator on the left-hand side of~(\ref{gPH}) must be positive-semidefinite.  Violation of the inequality is therefore a sufficient condition for the state to be nonseparable.~\footnote{In fact, if the state $\hat{\rho}_{1,2}$ is Gaussian, violation of inequality~(\ref{gPH}) is also necessary for the 
nonseparability of the state.}  It can be written in an equivalent form $\mathcal{P} \geq 0$, where $\mathcal{P}$ is a scalar function of the elements of the covariance matrix~\cite{Simon}.  Much of the literature works directly with the 
function $\mathcal{P}$, 
see e.g. Figure 4 of \cite{Finazzi-Carusotto-2014}.  This has the advantage of generality, but at the expense of obtuseness of the expressions.

If we restrict ourselves to 
states for which the two-point function $G^{(2)}(t,x;t',x')$ is either homogeneous (when the background is homogeneous) or stationary (when the background is stationary),
the gPH criterion can be simplified. 
%
Expectation values such as $\left\langle \hat{b}^{2}_{i} \right\rangle$ or $\left\langle \hat{b}^{\dagger}_{i} \hat{b}_{j} \right\rangle$ (where $i \neq j$) must then vanish, 
and the gPH criterion 
for nonseparability 
takes the 
form (see Appendix~B of~\cite{Adamek-Busch-Parentani}). 
\begin{equation}
\mathcal{P} = \left( \left(n_{1}+1\right) \left(n_{2}+1\right) - \left|c_{12}\right|^{2} \right) \left( n_{1} n_{2} - \left|c_{12}\right|^{2} \right) < 0 \,,
\label{gPH_hom}
\end{equation}
where $n_{i} = \left\langle \hat{b}_{i}^{\dagger} \hat{b}_{i} \right\rangle$ and $c_{12} = \left\langle \hat{b}_{1} \hat{b}_{2} \right\rangle$.
Recalling that for all states one has 
\begin{eqnarray} 
\left|c_{12}\right|^{2} &\leq& n_{1} \left(n_{2}+1\right) \nonumber\\
&\leq& n_{2}\left(n_{1}+1\right)\, , 
\label{maxim}
\end{eqnarray}
the first term in brackets in (\ref{gPH_hom}) is necessarily positive.  For this class of states, then, the gPH criterion is equivalent to
\begin{equation}
n_{1} n_{2} - \left|c_{12}\right|^{2} < 0 \,,
\label{gPH_hom2}
\end{equation}
which is exactly condition~(\ref{nonsep}).
Interestingly, for all two-mode states (not just the homogeneous/stationary 
subclass here considered), (\ref{gPH_hom2}) is sufficient for the gPH criterion to be satisfied, and hence for the state to be nonseparable (see Table 1 of~\cite{deNova-Sols-Zapata}, and Appendix B of~\cite{Adamek-Busch-Parentani}).

When assessing nonseparability 
in the homogeneous case (see Sec.~\ref{sub:HomoEntanglement}), we use another sufficient criterion, namely 
inequality~(\ref{strong_nonsep}). 
Given the form taken by $G^{(2)}(k,t)$ in a 
homogeneous state (see Eq.~(\ref{density-density})), this is equivalent to
\begin{equation}
\frac{n_{1}+n_{2}}{2} - \left|c_{12}\right| < 0 \, .
\label{strong_nonsep-prime}
\end{equation}
That is, the `product' condition~(\ref{gPH_hom2}) is replaced by the `sum' condition~(\ref{strong_nonsep-prime}).
In the isotropic case $n_{1}=n_{2}$, the proof of the sufficiency of the sum condition is immediate, for it is then equivalent to the product condition;
whereas in the anisotropic case $n_{1} \neq n_{2}$, the inequality 
\begin{equation}
\left(\frac{n_{1} + n_{2}}{2}\right)^2 - n_{1} n_{2} = \left(\frac{n_{1} - n_{2}}{2}\right)^2 > 0 
\end{equation}
guarantees that \eqref{strong_nonsep-prime} implies \eqref{gPH_hom2}.
The sum condition~(\ref{strong_nonsep-prime}) is thus a sufficient 
condition for nonseparability that can be accessed directly via measurements of 
$G^{(2)}(k,t)$ of Eq.~(\ref{density-density}). 

\subsubsection{Steerability}

For the subclass of homogeneous or stationary 
states, using Ref.~\cite{Jones-Wiseman-Doherty-PRA} a sufficient 
condition for 
steerability can be shown to be 
\begin{equation}
\Delta_{\rm steer}^{1 \to 2} \equiv n_{2} \left(n_{1}+\frac{1}{2}\right) - \left| c_{12} \right|^{2} < 0 \,.
\label{steer}
\end{equation}
Note that it is asymmetric with respect to the exchange 
of $n_{1}$ and $n_{2}$, which reflects the fact that it depends on which subsystem is being steered by the other. 
In the present case, it is subsystem 2 which is steered by subsystem 1,
hence the arrow in the superscript of $\Delta_{\rm steer}^{1 \to 2}$.  
Note 
that for Gaussian states, \eqref{steer} is also a necessary criterion. 

\begin{figure}
\includegraphics[width=0.45\columnwidth]{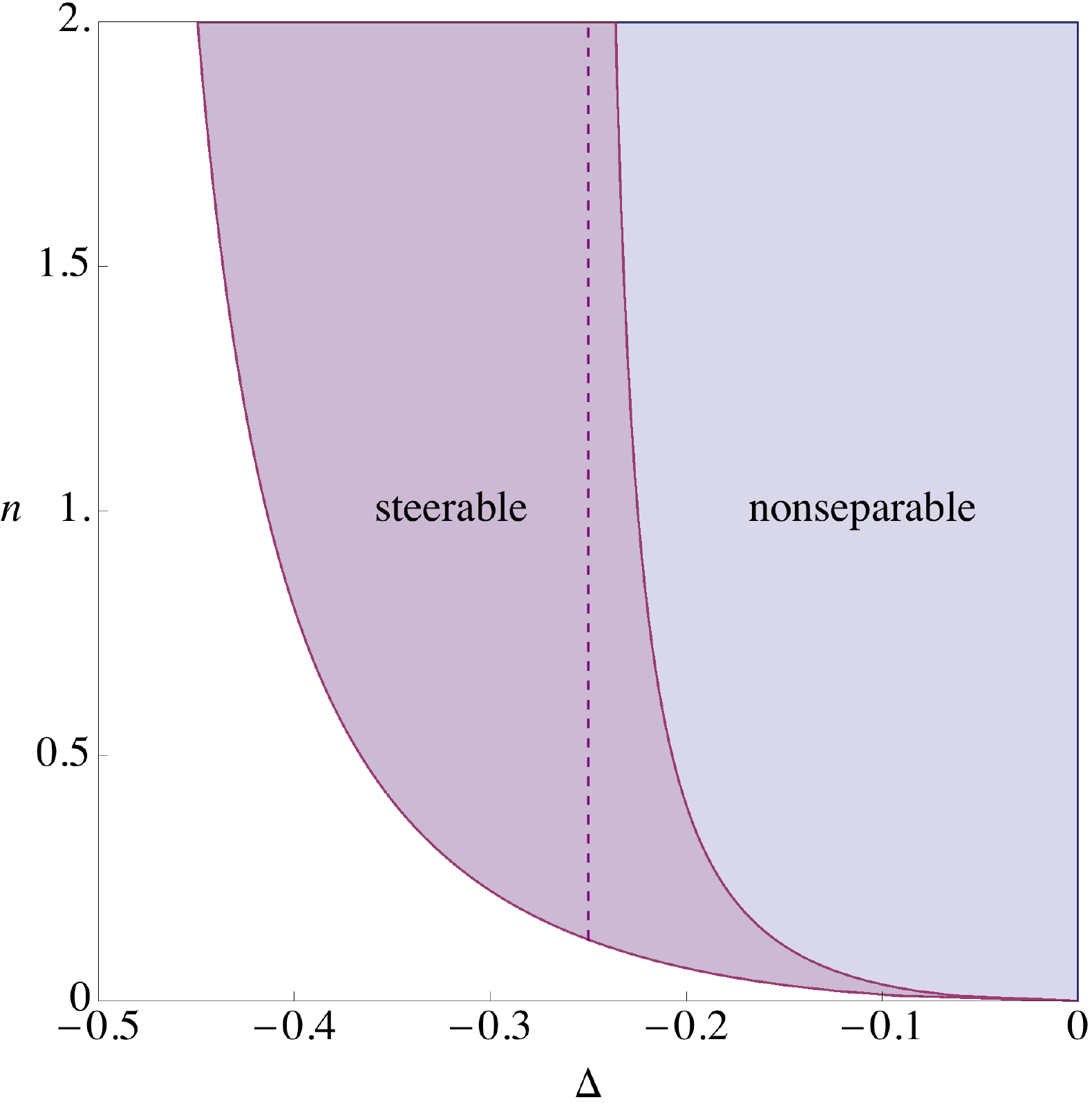}
\caption{Entangled states.  Assuming 
either homogeneity or stationarity of the two-mode state, along with isotropy $n_{1} = n_{2} \equiv n$, the color 
in the $(n,\Delta)$-plane (where $\Delta \equiv n - \left|c\right|$) determines the {\it minimum} 
degree of entanglement of the state. 
The white region is physically inaccessible, since it does not conform to 
\eqref{maxim}.  The entire shaded region corresponds to physical states for which $\Delta < 0$, and thus to nonseparable states; since we have assumed isotropy, the sum and product conditions (\ref{gPH_hom2}) and (\ref{strong_nonsep-prime}) are equivalent. 
Within this region, the darker shaded region corresponds to physical states for which 
(\ref{steer}) is satisfied. 
The dashed line corresponds to the strong steerability condition 
(\ref{strong_steer}). 
\label{fig:nonsep_steering}}
\end{figure}

As for nonseparability, there is a sum condition for steerability that is sufficient for 
inequality~(\ref{steer}) to be satisfied. 
To derive it, we use the inequality
\begin{equation}
\left(\frac{n_{1}+n_{2}}{2} + \frac{1}{4}\right)^{2} - n_{2}\left(n_{1}+\frac{1}{2}\right) = \left( \frac{n_{1}-n_{2}}{2} + \frac{1}{4} \right)^{2} \geq 0 \,, 
\end{equation}
which guarantees that (\ref{steer}) is satisfied whenever
\begin{equation}
\frac{n_{1}+n_{2}}{2} - \left|c_{12}\right| < -\frac{1}{4} \,.
\label{strong_steer}
\end{equation}
Unlike the product condition, the sum condition is symmetric under the exchange of $n_{1}$ and $n_{2}$; thus, if it is satisfied, each of the two subsystems is steerable by the other.
Notice that it has the same structure as the symmetric inequality~(\ref{strong_nonsep-prime}). 
Therefore, using Eq.~(\ref{density-density}), it can also be expressed directly in terms of $G^{(2)}(k,t)$ and $G^{(2)}_{\rm vac}(k)$, giving rise to inequality~(\ref{suff_steer}). 
This condition is represented by the dashed vertical line in Fig.~\ref{fig:nonsep_steering}
and by the horizontal dotted line in Fig.~\ref{fig:G2k_nonsep}.  

It should be noticed that it crosses the outermost limit of physical states with $n_{1} =n_{2} \equiv n$ for a finite value of the mean occupation number $n$.  
This can be understood from the fact that 
the realisation of inequality~(\ref{suff_steer}) 
requires sufficiently large oscillations of the $G^{(2)}(k,t)$ with respect to $G^{(2)}_{\rm vac}(k)$. 
It is interesting to notice that it is the symmetric 
condition~(\ref{strong_steer}) that is 
used in the recent work~\cite{Peise-et-al-2015}, where the threshold is correctly pointed out. 
Indeed, 
they measure the variances of 
linear combinations of operators pertaining to subsystems 1 and 2 (as in Eq.~(\ref{densityFT})), which in the language of Sec.~\ref{sec:Homogeneous} amounts to measuring $G^{(2)}(k,t)$ of Eq.~(\ref{density-density}). 

To summarize, 
although steerability is 
originally formulated in terms of inferred variances (see Eq.~(\ref{inf-dev})), 
a sufficient criterion 
can 
be expressed in terms of variances of linear combinations of operators pertaining to the two subsystems.  
Using the law of total variance~\cite{Ross}, this possibility follows from the inequality
\begin{equation}
\left(\Delta_{\rm inf}A_{2}\right)^{2} \leq \left\langle \left(\hat{A}_{1} \pm \hat{A}_{2} - \left\langle \hat{A}_{1} \pm \hat{A}_{2} \right\rangle \right)^{2} \right\rangle \,,
\end{equation}
which 
guarantees the sufficiency of Eq.~(\ref{strong_steer}) and therefore of Eq.~(\ref{suff_steer}).

For the interested reader, we add a simple illustration of the role of the inferred variance using 
the Wigner function of a Gaussian isotropic ($n_{1}=n_{2} \equiv n$) 
state (for a similar analysis based on the Husimi $Q$-distribution, see Appendix~D of \cite{Campo-Parentani-2005}). 
The state is thus completely characterized by the expectation values $n$ and $c_{12}$.  We further assume that the correlation term $c_{12}$ 
is real, 
so that 
the quadrature operators $\hat{q}_{j}$ and $\hat{p}_{j}$ introduced above can be used 
in 
the steerability criterion~(\ref{steering_defn}). 
Using as variables the coherent state amplitudes $u_{j} = \left( q_{j} + i p_{j} \right)/\sqrt{2}$, the Wigner function of the bipartite state $\hat{\rho}_{1,2}$ is given 
by
\begin{equation}
W_{1,2}\left(u_{1},u_{2}\right) = \mathcal{N} \, \mathrm{exp}\left(-\frac{\left|u_{1}\right|^{2}}{n+1/2}\right) \, \mathrm{exp}\left(-\frac{\left|u_{2}-\bar{u}_{2}\left(u_{1}\right)\right|^{2}}{\left(\Delta_{\rm inf}q_{2}\right)^{2}}\right) \,,
\label{wigner}
\end{equation}
where $\mathcal{N}$ is a normalization prefactor, and where we have defined
\begin{equation}
\bar{u}_{2}\left(u_{1}\right) = \frac{c_{12}}{n+1/2} \, u_{1}^{\star} \,, \qquad \left(\Delta_{\rm inf}q_{2}\right)^{2} = \frac{1}{2} + \frac{\Delta_{\rm steer}^{1 \to 2}}{n+1/2} \,. 
\label{wigner_defns}
\end{equation}
Straightforward symmetry arguments show that 
$\left(\Delta_{\rm inf}p_{2}\right)^{2} = \left(\Delta_{\rm inf}q_{2}\right)^{2}$. The first exponential factor in Eq.~(\ref{wigner}) is (up to a normalization prefactor) the reduced Wigner function of subsystem 1 
having traced over the degrees of freedom pertaining to subsystem 2.   It is characterized by a width which is the total standard deviation of $\hat{q}_{1}$ (or $\hat{p}_{1}$).  The second factor is the 
conditional Wigner function of subsystem 2 given that a measurement of $\hat{b}_{1}$ 
yields the value $u_{1}$.  It is characterized both by a conditional mean $\bar{u}_{2}\left(u_{1}\right)$ and by a width which is the inferred standard deviation of $\hat{q}_{2}$ (or $\hat{p}_{2}$) given a measurement of $\hat{q}_{1}$ (or $\hat{p}_{1}$).  Subsystem 2 is steerable by subsystem 1 whenever the inferred variance 
is smaller than its vacuum value of $1/2$, 
which occurs precisely when
condition~(\ref{steer}) is satisfied.



\section{Additional information from other types of measurement
\label{app:OtherMeasurements}}

\subsection{
Phase fluctuations and non-commuting measurements}

In addition to density fluctuations, it is interesting to study 
the two-point correlation functions 
involving the phase fluctuations $\delta \theta$ 
in order to see what is the extra information about $n_k$ and $c_k$ that could be extracted. 
Using again $\hat{\Phi}(t,x) = e^{-i \mu t + i K x} \left( \Phi_{0}
 + \delta\hat{\phi}(t,x) \right)$, one has
\begin{equation}
\delta\hat{\phi}(t,x) = 
\frac{\delta \hat{\rho}}{2 \sqrt{\rho_{0}}} + i \,\sqrt{\rho_{0}} \, \delta \hat{\theta} \,. 
\end{equation}
Then, as in \eqref{densityFTphonon}, 
 it is useful to work with the spatial Fourier transform and to express it 
using the phonon operators. One finds 
\begin{equation}
\delta \hat{\theta}_{k} = \frac{\sqrt{N}}{2 i \rho_{0}} 
\left( u_{k} - v_{k} \right) \left( \hat{\varphi}_{k} - \hat{\varphi}_{-k}^{\dagger} \right) \,.
\label{densityFTphase}
\end{equation}
Much as for the $\hat{\rho}_{k}$, we here have $\hat{\theta}_{k}^{\dagger} = \hat{\theta}_{-k}$ and $\left[ \hat{\theta}_{k}, \, \hat{\theta}_{k}^{\dagger} \right] = 0$. Therefore, the correlation $\left\langle \left| \hat{\theta}_{k} \right|^{2} \right\rangle$ is well-defined, and we have
\begin{eqnarray}
\left\langle \left| \hat{\theta}_{k} \right|^{2} \right\rangle & = & \left\langle \hat{\theta}_{k} \hat{\theta}_{-k} \right\rangle \nonumber \\ & = & \frac{N}{4\rho_{0}^{2}} 
\left( u_{k}-v_{k} \right)^{2} \left( 1 + n_{k} + n_{-k} - 2 \, \mathrm{Re} \left[ c_{k} e^{-2 i \omega_{k} t} \right] \right) \,.
\label{phase-phase}
\end{eqnarray}
Comparing with \eq{density-density}, we see that we gain little additional information from phase measurements: indeed, apart from the minus sign in the prefactor $\left(u_{k}-v_{k}\right)^{2}$, the essential difference occurs in the minus sign in front of the oscillating term.  Thus, the measurement is just as if we had examined the density-density correlation shifted in time by half a period.  

To conclude, 
let us now consider the information encoded in density-phase correlations:
\begin{equation}
\left\langle \left\{ \hat{\rho}_{k}, \, \hat{\theta}_{-k} \right\} \right\rangle = 2\,L\,\left( i\, \delta n_{k} + \mathrm{Im}\left[ c_{k} e^{-2 i \omega_{k} t} \right] \right) \,, 
\label{density-phase}
\end{equation}
where curly brackets signify the anti-commutator (used here because $\hat{\rho}_{k}$ and $\hat{\theta}_{-k}$ do {\it not} commute, since they obey $[\hat{\rho}_{k},\hat{\theta}_{-k'}]= i\,L\,\delta_{k,k'}$). 
This measurement {\it does} give us access to new information: the degree of anisotropy $\delta n_{k} = \left(n_k - n_{-k}\right)/2$. 

\subsection{Anisotropy}

It is useful to investigate the effects of anisotropy further.  Let us define $\delta n_{k} \equiv \left(n_{k} - n_{-k}\right)/2$. 
Then the observable accessible from measurements of $\left\langle \hat{\rho}_{k} \hat{\rho}_{-k} \right\rangle$ is
\begin{equation}
\frac{1}{2} \left(n_{k} + n_{-k} \right) - \left| c_{k} \right| = n_{k} - \left| c_{k} \right| - \delta n_{k} \,. 
\end{equation}
On the other hand, the criterion that determines separability is $n_{k} n_{-k} - \left| c_{k} \right|^{2} > 0$ or, taking the square root for ease of comparison with measurements, $ \sqrt{n_{k} n_{-k}} - \left| c_{k} \right| > 0$.  Substituting $\delta n_{k}$ and Taylor expanding the square root, we find
\begin{eqnarray}
\sqrt{n_{k} n_{-k}} - \left| c_{k} \right| & = & n_{k} \sqrt{1 -2\, \delta n_{k} / n_{k}} - \left| c_{k} \right| \nonumber \\ 
& = & n_{k} - \left| c_{k} \right| - \delta n_{k} - \frac{\left(\delta n_{k}\right)^{2}}{2 n_{k}} + O\left(\left(\delta n_{k}\right)^{3}\right) \,. 
\end{eqnarray}
Therefore, the theoretical and measurable criteria first differ at quadratic order in $\delta n_{k}$.  That is, if there is a small degree of anisotropy, the theoretical separability threshold occurs slightly above that directly accessible to measurement: there is a small band just above the vacuum value of $\left\langle \hat{\rho}_{k} \hat{\rho}_{-k} \right\rangle$ where the corresponding state is nonseparable, and its thickness varies as $\left(\delta n_{k} \right)^{2}$.

Continuing in this vein, let us assume that we can place a lower bound on the degree of anisotropy, i.e. we have $\left| \delta n_{k} \right| > M_{k} > 0$ for some $M_{k}$.  From what has been said so far, it is clear that there will exist states which do not satisfy $\frac{1}{2}\left(n_{k}+n_{-k}\right) - \left|c_{k}\right| < 0$, but which are nonetheless nonseparable.  Can we improve the sufficiency criterion in order to be able to recognise some of these states?  
We shall allow ourselves to collect measurements of $G^{(2)}(k,t)$ at different times, but even then, it is clear from Eq.~(\ref{density-density}) 
that we have experimental access only to $n_{k}+n_{-k}$ and $\left|c_{k}\right|$. 
The improved criterion must therefore 
contain only these values (in addition to the bound $M_{k}$).  Let us first assume separability, i.e. $n_{k} n_{-k} - \left| c_{k} \right|^{2} \geq 0$, and derive a necessary condition for this to be true; any violation of this condition will then constitute a sufficient condition for nonseparability.  Firstly, we note that an equivalent separability criterion (from straightforward algebraic rearrangement) is
\begin{equation}
2 \left( \delta n_{k} \right)^{2} \leq \frac{1}{2}\left(n_{k}^{2} + n_{-k}^{2}\right) - \left|c_{k}\right|^{2} \,. 
\label{sep_anisotropy}
\end{equation}
We then have
\begin{eqnarray}
4 \left(\delta n_{k}\right)^{2} & \leq & 4 \left(\delta n_{k}\right)^{2} + 2 \left( n_{k} n_{-k} - \left| c_{k} \right|^{2} \right) \nonumber \\ 
& \leq & \left( n_{k} + n_{-k} \right)^{2} - 4 \left| c_{k} \right|^{2} \nonumber \\
& = & \left( n_{k} + n_{-k} - 2 \left|c_{k}\right| \right) \left( n_{k} + n_{-k} + 2 \left|c_{k}\right| \right) \,,
\end{eqnarray}
where in the first line we have used the standard criterion for separability and in the second line we have used the equivalent condition (\ref{sep_anisotropy}).  The result can be rewritten in the form
\begin{equation}
\frac{1}{2}\left(n_{k}+n_{-k}\right) - \left|c_{k}\right| \geq \frac{2 \left(\delta n_{k}\right)^{2}}{n_{k}+n_{-k}+2\left|c_{k}\right|} \,. 
\end{equation}
Violation of this inequality is therefore sufficient to be able to assert the nonseparability of the state, and if we assume a known lower bound for $\left| \delta n_{k} \right| > M_{k}$, we can shift the nonseparability criterion to
\begin{equation}
\frac{1}{2}\left( n_{k} + n_{-k} \right) - \left| c_{k} \right| < \frac{2 M_{k}^{2}}{n_{k}+n_{-k}+2\left|c_{k}\right|} \,. 
\end{equation}

\begin{figure}
\includegraphics[width=0.45\columnwidth]{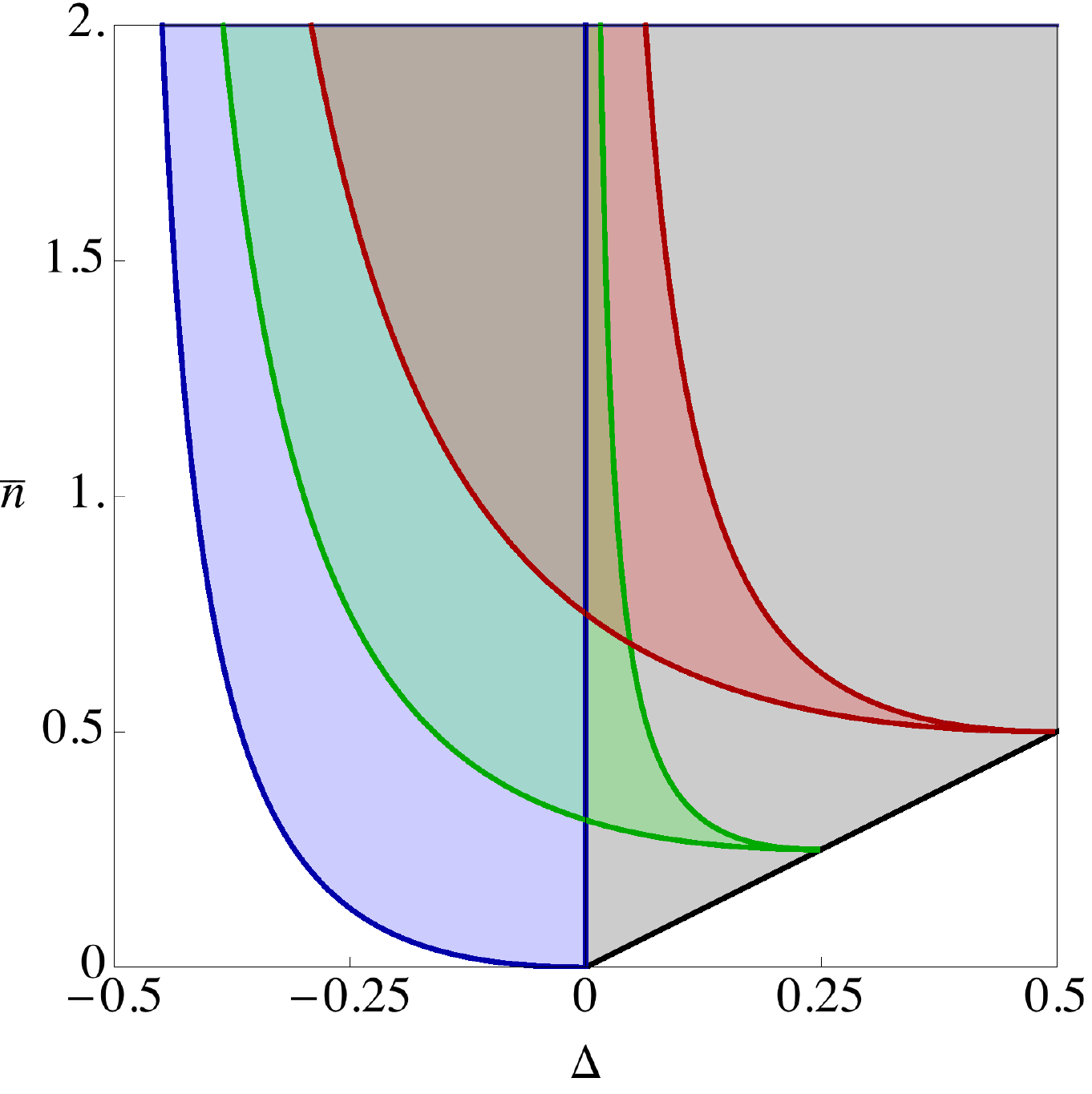}
\caption{Nonseparable states.  Assuming 
either homogeneity or stationarity of the two-mode state, and if the degree of anistropy $\delta n \equiv \left|n_{1} - n_{2}\right|/2$ is known, the point in the $(\bar{n}, \Delta)$-plane (where $\bar{n} \equiv (n_{1}+n_{2})/2$ and $\Delta \equiv \bar{n} - \left|c\right|$) determines the {\it minimum} degree of nonseparability of the state (the equality being reached for Gaussian states). 
The white region is 
inaccessible to quantum mechanical states. 
Within the shaded region, we have used different colors to illustrate the set of all nonseparable states when $\delta n = 0$ (blue), $1/4$ (green) and $1/2$ (red).  
The gray regions correspond to separable states. 
The left-most colored curve always represents the boundary of quantum mechanical states given the value of $\delta n$. 
\label{fig:nonsep_aniso}}
\end{figure}

The knowledge of $\delta n$ on nonseparability is clearly illustrated in Figure~\ref{fig:nonsep_aniso}.

\end{appendices}

\bibliography{biblio}

\begin{thebibliography}{47}%
\makeatletter
\providecommand \@ifxundefined [1]{%
 \@ifx{#1\undefined}
}%
\providecommand \@ifnum [1]{%
 \ifnum #1\expandafter \@firstoftwo
 \else \expandafter \@secondoftwo
 \fi
}%
\providecommand \@ifx [1]{%
 \ifx #1\expandafter \@firstoftwo
 \else \expandafter \@secondoftwo
 \fi
}%
\providecommand \natexlab [1]{#1}%
\providecommand \enquote  [1]{``#1''}%
\providecommand \bibnamefont  [1]{#1}%
\providecommand \bibfnamefont [1]{#1}%
\providecommand \citenamefont [1]{#1}%
\providecommand \href@noop [0]{\@secondoftwo}%
\providecommand \href [0]{\begingroup \@sanitize@url \@href}%
\providecommand \@href[1]{\@@startlink{#1}\@@href}%
\providecommand \@@href[1]{\endgroup#1\@@endlink}%
\providecommand \@sanitize@url [0]{\catcode `\\12\catcode `\$12\catcode
  `\&12\catcode `\#12\catcode `\^12\catcode `\_12\catcode `\%12\relax}%
\providecommand \@@startlink[1]{}%
\providecommand \@@endlink[0]{}%
\providecommand \url  [0]{\begingroup\@sanitize@url \@url }%
\providecommand \@url [1]{\endgroup\@href {#1}{\urlprefix }}%
\providecommand \urlprefix  [0]{URL }%
\providecommand \Eprint [0]{\href }%
\providecommand \doibase [0]{http://dx.doi.org/}%
\providecommand \selectlanguage [0]{\@gobble}%
\providecommand \bibinfo  [0]{\@secondoftwo}%
\providecommand \bibfield  [0]{\@secondoftwo}%
\providecommand \translation [1]{[#1]}%
\providecommand \BibitemOpen [0]{}%
\providecommand \bibitemStop [0]{}%
\providecommand \bibitemNoStop [0]{.\EOS\space}%
\providecommand \EOS [0]{\spacefactor3000\relax}%
\providecommand \BibitemShut  [1]{\csname bibitem#1\endcsname}%
\let\auto@bib@innerbib\@empty
\bibitem [{\citenamefont {Birrell}\ and\ \citenamefont
  {Davies}(1982)}]{Birrell-Davies}%
  \BibitemOpen
  \bibfield  {author} {\bibinfo {author} {\bibfnamefont {N.~D.}\ \bibnamefont
  {Birrell}}\ and\ \bibinfo {author} {\bibfnamefont {P.~C.~W.}\ \bibnamefont
  {Davies}},\ }\href@noop {} {\emph {\bibinfo {title} {Quantum fields in curved
  space}}}\ (\bibinfo  {publisher} {Cambridge University Press},\ \bibinfo
  {year} {1982})\BibitemShut {NoStop}%
\bibitem [{\citenamefont {Carusotto}\ \emph {et~al.}(2010)\citenamefont
  {Carusotto}, \citenamefont {Balbinot}, \citenamefont {Fabbri},\ and\
  \citenamefont {Recati}}]{Carusotto-DCE}%
  \BibitemOpen
  \bibfield  {author} {\bibinfo {author} {\bibfnamefont {I.}~\bibnamefont
  {Carusotto}}, \bibinfo {author} {\bibfnamefont {R.}~\bibnamefont {Balbinot}},
  \bibinfo {author} {\bibfnamefont {A.}~\bibnamefont {Fabbri}}, \ and\ \bibinfo
  {author} {\bibfnamefont {A.}~\bibnamefont {Recati}},\ }\href {\doibase
  10.1140/epjd/e2009-00314-3} {\bibfield  {journal} {\bibinfo  {journal} {The
  European Physical Journal D}\ }\textbf {\bibinfo {volume} {56}},\ \bibinfo
  {pages} {391} (\bibinfo {year} {2010})}\BibitemShut {NoStop}%
\bibitem [{\citenamefont {Jaskula}\ \emph {et~al.}(2012)\citenamefont
  {Jaskula}, \citenamefont {Partridge}, \citenamefont {Bonneau}, \citenamefont
  {Lopes}, \citenamefont {Ruaudel}, \citenamefont {Boiron},\ and\ \citenamefont
  {Westbrook}}]{Jaskula-et-al}%
  \BibitemOpen
  \bibfield  {author} {\bibinfo {author} {\bibfnamefont {J.-C.}\ \bibnamefont
  {Jaskula}}, \bibinfo {author} {\bibfnamefont {G.~B.}\ \bibnamefont
  {Partridge}}, \bibinfo {author} {\bibfnamefont {M.}~\bibnamefont {Bonneau}},
  \bibinfo {author} {\bibfnamefont {R.}~\bibnamefont {Lopes}}, \bibinfo
  {author} {\bibfnamefont {J.}~\bibnamefont {Ruaudel}}, \bibinfo {author}
  {\bibfnamefont {D.}~\bibnamefont {Boiron}}, \ and\ \bibinfo {author}
  {\bibfnamefont {C.~I.}\ \bibnamefont {Westbrook}},\ }\href {\doibase
  10.1103/PhysRevLett.109.220401} {\bibfield  {journal} {\bibinfo  {journal}
  {Phys. Rev. Lett.}\ }\textbf {\bibinfo {volume} {109}},\ \bibinfo {pages}
  {220401} (\bibinfo {year} {2012})}\BibitemShut {NoStop}%
\bibitem [{\citenamefont {Robertson}\ \emph {et~al.}(2017)\citenamefont
  {Robertson}, \citenamefont {Michel},\ and\ \citenamefont
  {Parentani}}]{Robertson-Michel-Parentani-2017}%
  \BibitemOpen
  \bibfield  {author} {\bibinfo {author} {\bibfnamefont {S.}~\bibnamefont
  {Robertson}}, \bibinfo {author} {\bibfnamefont {F.}~\bibnamefont {Michel}}, \
  and\ \bibinfo {author} {\bibfnamefont {R.}~\bibnamefont {Parentani}},\ }\href
  {\doibase 10.1103/PhysRevD.95.065020} {\bibfield  {journal} {\bibinfo
  {journal} {Phys. Rev. D}\ }\textbf {\bibinfo {volume} {95}},\ \bibinfo
  {pages} {065020} (\bibinfo {year} {2017})}\BibitemShut {NoStop}%
\bibitem [{\citenamefont {Campo}\ and\ \citenamefont
  {Parentani}(2004)}]{Campo-Parentani-2004}%
  \BibitemOpen
  \bibfield  {author} {\bibinfo {author} {\bibfnamefont {D.}~\bibnamefont
  {Campo}}\ and\ \bibinfo {author} {\bibfnamefont {R.}~\bibnamefont
  {Parentani}},\ }\href@noop {} {\bibfield  {journal} {\bibinfo  {journal}
  {Phys. Rev. D}\ }\textbf {\bibinfo {volume} {70}},\ \bibinfo {pages} {105020}
  (\bibinfo {year} {2004})}\BibitemShut {NoStop}%
\bibitem [{\citenamefont {Brout}\ \emph {et~al.}(1995)\citenamefont {Brout},
  \citenamefont {Massar}, \citenamefont {Parentani},\ and\ \citenamefont
  {Spindel}}]{Primer}%
  \BibitemOpen
  \bibfield  {author} {\bibinfo {author} {\bibfnamefont {R.}~\bibnamefont
  {Brout}}, \bibinfo {author} {\bibfnamefont {S.}~\bibnamefont {Massar}},
  \bibinfo {author} {\bibfnamefont {R.}~\bibnamefont {Parentani}}, \ and\
  \bibinfo {author} {\bibfnamefont {P.}~\bibnamefont {Spindel}},\ }\href
  {\doibase http://dx.doi.org/10.1016/0370-1573(95)00008-5} {\bibfield
  {journal} {\bibinfo  {journal} {Physics Reports}\ }\textbf {\bibinfo {volume}
  {260}},\ \bibinfo {pages} {329 } (\bibinfo {year} {1995})}\BibitemShut
  {NoStop}%
\bibitem [{\citenamefont {Jones}\ \emph {et~al.}(2007)\citenamefont {Jones},
  \citenamefont {Wiseman},\ and\ \citenamefont
  {Doherty}}]{Jones-Wiseman-Doherty-PRA}%
  \BibitemOpen
  \bibfield  {author} {\bibinfo {author} {\bibfnamefont {S.~J.}\ \bibnamefont
  {Jones}}, \bibinfo {author} {\bibfnamefont {H.~M.}\ \bibnamefont {Wiseman}},
  \ and\ \bibinfo {author} {\bibfnamefont {A.~C.}\ \bibnamefont {Doherty}},\
  }\href {\doibase 10.1103/PhysRevA.76.052116} {\bibfield  {journal} {\bibinfo
  {journal} {Phys. Rev. A}\ }\textbf {\bibinfo {volume} {76}},\ \bibinfo
  {pages} {052116} (\bibinfo {year} {2007})}\BibitemShut {NoStop}%
\bibitem [{\citenamefont {Werner}(1989)}]{Werner}%
  \BibitemOpen
  \bibfield  {author} {\bibinfo {author} {\bibfnamefont {R.~F.}\ \bibnamefont
  {Werner}},\ }\href {\doibase 10.1103/PhysRevA.40.4277} {\bibfield  {journal}
  {\bibinfo  {journal} {Phys. Rev. A}\ }\textbf {\bibinfo {volume} {40}},\
  \bibinfo {pages} {4277} (\bibinfo {year} {1989})}\BibitemShut {NoStop}%
\bibitem [{\citenamefont {Simon}(2000)}]{Simon}%
  \BibitemOpen
  \bibfield  {author} {\bibinfo {author} {\bibfnamefont {R.}~\bibnamefont
  {Simon}},\ }\href {\doibase 10.1103/PhysRevLett.84.2726} {\bibfield
  {journal} {\bibinfo  {journal} {Phys. Rev. Lett.}\ }\textbf {\bibinfo
  {volume} {84}},\ \bibinfo {pages} {2726} (\bibinfo {year}
  {2000})}\BibitemShut {NoStop}%
\bibitem [{\citenamefont {Wiseman}\ \emph {et~al.}(2007)\citenamefont
  {Wiseman}, \citenamefont {Jones},\ and\ \citenamefont
  {Doherty}}]{Wiseman-Jones-Doherty-PRL}%
  \BibitemOpen
  \bibfield  {author} {\bibinfo {author} {\bibfnamefont {H.~M.}\ \bibnamefont
  {Wiseman}}, \bibinfo {author} {\bibfnamefont {S.~J.}\ \bibnamefont {Jones}},
  \ and\ \bibinfo {author} {\bibfnamefont {A.~C.}\ \bibnamefont {Doherty}},\
  }\href {\doibase 10.1103/PhysRevLett.98.140402} {\bibfield  {journal}
  {\bibinfo  {journal} {Phys. Rev. Lett.}\ }\textbf {\bibinfo {volume} {98}},\
  \bibinfo {pages} {140402} (\bibinfo {year} {2007})}\BibitemShut {NoStop}%
\bibitem [{\citenamefont {Campo}\ and\ \citenamefont
  {Parentani}(2005)}]{Campo-Parentani-2005}%
  \BibitemOpen
  \bibfield  {author} {\bibinfo {author} {\bibfnamefont {D.}~\bibnamefont
  {Campo}}\ and\ \bibinfo {author} {\bibfnamefont {R.}~\bibnamefont
  {Parentani}},\ }\href {\doibase 10.1103/PhysRevD.72.045015} {\bibfield
  {journal} {\bibinfo  {journal} {Phys. Rev. D}\ }\textbf {\bibinfo {volume}
  {72}},\ \bibinfo {pages} {045015} (\bibinfo {year} {2005})}\BibitemShut
  {NoStop}%
\bibitem [{\citenamefont {Adamek}\ \emph {et~al.}(2013)\citenamefont {Adamek},
  \citenamefont {Busch},\ and\ \citenamefont
  {Parentani}}]{Adamek-Busch-Parentani}%
  \BibitemOpen
  \bibfield  {author} {\bibinfo {author} {\bibfnamefont {J.}~\bibnamefont
  {Adamek}}, \bibinfo {author} {\bibfnamefont {X.}~\bibnamefont {Busch}}, \
  and\ \bibinfo {author} {\bibfnamefont {R.}~\bibnamefont {Parentani}},\ }\href
  {\doibase 10.1103/PhysRevD.87.124039} {\bibfield  {journal} {\bibinfo
  {journal} {Phys. Rev. D}\ }\textbf {\bibinfo {volume} {87}},\ \bibinfo
  {pages} {124039} (\bibinfo {year} {2013})}\BibitemShut {NoStop}%
\bibitem [{\citenamefont {de~Nova}\ \emph {et~al.}(2015)\citenamefont
  {de~Nova}, \citenamefont {Sols},\ and\ \citenamefont
  {Zapata}}]{deNova-Sols-Zapata}%
  \BibitemOpen
  \bibfield  {author} {\bibinfo {author} {\bibfnamefont {J.~R.~M.}\
  \bibnamefont {de~Nova}}, \bibinfo {author} {\bibfnamefont {F.}~\bibnamefont
  {Sols}}, \ and\ \bibinfo {author} {\bibfnamefont {I.}~\bibnamefont
  {Zapata}},\ }\href {http://stacks.iop.org/1367-2630/17/i=10/a=105003}
  {\bibfield  {journal} {\bibinfo  {journal} {New Journal of Physics}\ }\textbf
  {\bibinfo {volume} {17}},\ \bibinfo {pages} {105003} (\bibinfo {year}
  {2015})}\BibitemShut {NoStop}%
\bibitem [{\citenamefont {Banaszek}\ and\ \citenamefont
  {W\'odkiewicz}(1998)}]{Banaszek-1998}%
  \BibitemOpen
  \bibfield  {author} {\bibinfo {author} {\bibfnamefont {K.}~\bibnamefont
  {Banaszek}}\ and\ \bibinfo {author} {\bibfnamefont {K.}~\bibnamefont
  {W\'odkiewicz}},\ }\href {\doibase 10.1103/PhysRevA.58.4345} {\bibfield
  {journal} {\bibinfo  {journal} {Phys. Rev. A}\ }\textbf {\bibinfo {volume}
  {58}},\ \bibinfo {pages} {4345} (\bibinfo {year} {1998})}\BibitemShut
  {NoStop}%
\bibitem [{\citenamefont {Campo}\ and\ \citenamefont
  {Parentani}(2006)}]{Campo-Parentani-2006}%
  \BibitemOpen
  \bibfield  {author} {\bibinfo {author} {\bibfnamefont {D.}~\bibnamefont
  {Campo}}\ and\ \bibinfo {author} {\bibfnamefont {R.}~\bibnamefont
  {Parentani}},\ }\href {\doibase 10.1103/PhysRevD.74.025001} {\bibfield
  {journal} {\bibinfo  {journal} {Phys. Rev. D}\ }\textbf {\bibinfo {volume}
  {74}},\ \bibinfo {pages} {025001} (\bibinfo {year} {2006})}\BibitemShut
  {NoStop}%
\bibitem [{\citenamefont {Finazzi}\ and\ \citenamefont
  {Carusotto}(2014)}]{Finazzi-Carusotto-2014}%
  \BibitemOpen
  \bibfield  {author} {\bibinfo {author} {\bibfnamefont {S.}~\bibnamefont
  {Finazzi}}\ and\ \bibinfo {author} {\bibfnamefont {I.}~\bibnamefont
  {Carusotto}},\ }\href {\doibase 10.1103/PhysRevA.90.033607} {\bibfield
  {journal} {\bibinfo  {journal} {Phys. Rev. A}\ }\textbf {\bibinfo {volume}
  {90}},\ \bibinfo {pages} {033607} (\bibinfo {year} {2014})}\BibitemShut
  {NoStop}%
\bibitem [{\citenamefont {Finke}\ \emph {et~al.}(2016)\citenamefont {Finke},
  \citenamefont {Jain},\ and\ \citenamefont
  {Weinfurtner}}]{Finke-Jain-Weinfurtner}%
  \BibitemOpen
  \bibfield  {author} {\bibinfo {author} {\bibfnamefont {A.}~\bibnamefont
  {Finke}}, \bibinfo {author} {\bibfnamefont {P.}~\bibnamefont {Jain}}, \ and\
  \bibinfo {author} {\bibfnamefont {S.}~\bibnamefont {Weinfurtner}},\ }\href
  {http://stacks.iop.org/1367-2630/18/i=11/a=113017} {\bibfield  {journal}
  {\bibinfo  {journal} {New Journal of Physics}\ }\textbf {\bibinfo {volume}
  {18}},\ \bibinfo {pages} {113017} (\bibinfo {year} {2016})}\BibitemShut
  {NoStop}%
\bibitem [{\citenamefont {Busch}\ and\ \citenamefont
  {Parentani}(2013)}]{Busch-Parentani-2013}%
  \BibitemOpen
  \bibfield  {author} {\bibinfo {author} {\bibfnamefont {X.}~\bibnamefont
  {Busch}}\ and\ \bibinfo {author} {\bibfnamefont {R.}~\bibnamefont
  {Parentani}},\ }\href {\doibase 10.1103/PhysRevD.88.045023} {\bibfield
  {journal} {\bibinfo  {journal} {Phys. Rev. D}\ }\textbf {\bibinfo {volume}
  {88}},\ \bibinfo {pages} {045023} (\bibinfo {year} {2013})}\BibitemShut
  {NoStop}%
\bibitem [{\citenamefont {Busch}\ \emph {et~al.}(2014)\citenamefont {Busch},
  \citenamefont {Carusotto},\ and\ \citenamefont
  {Parentani}}]{Busch-Carusotto-Parentani}%
  \BibitemOpen
  \bibfield  {author} {\bibinfo {author} {\bibfnamefont {X.}~\bibnamefont
  {Busch}}, \bibinfo {author} {\bibfnamefont {I.}~\bibnamefont {Carusotto}}, \
  and\ \bibinfo {author} {\bibfnamefont {R.}~\bibnamefont {Parentani}},\ }\href
  {\doibase 10.1103/PhysRevA.89.043819} {\bibfield  {journal} {\bibinfo
  {journal} {Phys. Rev. A}\ }\textbf {\bibinfo {volume} {89}},\ \bibinfo
  {pages} {043819} (\bibinfo {year} {2014})}\BibitemShut {NoStop}%
\bibitem [{\citenamefont {Steinhauer}(2015)}]{Steinhauer-2015}%
  \BibitemOpen
  \bibfield  {author} {\bibinfo {author} {\bibfnamefont {J.}~\bibnamefont
  {Steinhauer}},\ }\href {\doibase 10.1103/PhysRevD.92.024043} {\bibfield
  {journal} {\bibinfo  {journal} {Phys. Rev. D}\ }\textbf {\bibinfo {volume}
  {92}},\ \bibinfo {pages} {024043} (\bibinfo {year} {2015})}\BibitemShut
  {NoStop}%
\bibitem [{\citenamefont {Steinhauer}(2016)}]{Steinhauer-2016}%
  \BibitemOpen
  \bibfield  {author} {\bibinfo {author} {\bibfnamefont {J.}~\bibnamefont
  {Steinhauer}},\ }\href@noop {} {\bibfield  {journal} {\bibinfo  {journal}
  {Nat. Phys.}\ }\textbf {\bibinfo {volume} {12}},\ \bibinfo {pages} {959}
  (\bibinfo {year} {2016})}\BibitemShut {NoStop}%
\bibitem [{\citenamefont {Busch}\ and\ \citenamefont
  {Parentani}(2014)}]{Busch-Parentani-2014}%
  \BibitemOpen
  \bibfield  {author} {\bibinfo {author} {\bibfnamefont {X.}~\bibnamefont
  {Busch}}\ and\ \bibinfo {author} {\bibfnamefont {R.}~\bibnamefont
  {Parentani}},\ }\href {\doibase 10.1103/PhysRevD.89.105024} {\bibfield
  {journal} {\bibinfo  {journal} {Phys. Rev. D}\ }\textbf {\bibinfo {volume}
  {89}},\ \bibinfo {pages} {105024} (\bibinfo {year} {2014})}\BibitemShut
  {NoStop}%
\bibitem [{\citenamefont {Menotti}\ and\ \citenamefont
  {Stringari}(2002)}]{Menotti-Stringari}%
  \BibitemOpen
  \bibfield  {author} {\bibinfo {author} {\bibfnamefont {C.}~\bibnamefont
  {Menotti}}\ and\ \bibinfo {author} {\bibfnamefont {S.}~\bibnamefont
  {Stringari}},\ }\href {\doibase 10.1103/PhysRevA.66.043610} {\bibfield
  {journal} {\bibinfo  {journal} {Phys. Rev. A}\ }\textbf {\bibinfo {volume}
  {66}},\ \bibinfo {pages} {043610} (\bibinfo {year} {2002})}\BibitemShut
  {NoStop}%
\bibitem [{\citenamefont {Tozzo}\ and\ \citenamefont
  {Dalfovo}(2004)}]{Tozzo-Dalfovo}%
  \BibitemOpen
  \bibfield  {author} {\bibinfo {author} {\bibfnamefont {C.}~\bibnamefont
  {Tozzo}}\ and\ \bibinfo {author} {\bibfnamefont {F.}~\bibnamefont
  {Dalfovo}},\ }\href {\doibase 10.1103/PhysRevA.69.053606} {\bibfield
  {journal} {\bibinfo  {journal} {Phys. Rev. A}\ }\textbf {\bibinfo {volume}
  {69}},\ \bibinfo {pages} {053606} (\bibinfo {year} {2004})}\BibitemShut
  {NoStop}%
\bibitem [{\citenamefont {Gerbier}(2004)}]{Gerbier}%
  \BibitemOpen
  \bibfield  {author} {\bibinfo {author} {\bibfnamefont {F.}~\bibnamefont
  {Gerbier}},\ }\href@noop {} {\bibfield  {journal} {\bibinfo  {journal}
  {Europhys. Lett.}\ }\textbf {\bibinfo {volume} {66}},\ \bibinfo {pages} {771}
  (\bibinfo {year} {2004})}\BibitemShut {NoStop}%
\bibitem [{\citenamefont {Dalfovo}\ \emph {et~al.}(1999)\citenamefont
  {Dalfovo}, \citenamefont {Giorgini}, \citenamefont {Pitaevskii},\ and\
  \citenamefont {Stringari}}]{Dalfovo-et-al-1999}%
  \BibitemOpen
  \bibfield  {author} {\bibinfo {author} {\bibfnamefont {F.}~\bibnamefont
  {Dalfovo}}, \bibinfo {author} {\bibfnamefont {S.}~\bibnamefont {Giorgini}},
  \bibinfo {author} {\bibfnamefont {L.~P.}\ \bibnamefont {Pitaevskii}}, \ and\
  \bibinfo {author} {\bibfnamefont {S.}~\bibnamefont {Stringari}},\ }\href@noop
  {} {\bibfield  {journal} {\bibinfo  {journal} {Rev. Mod. Phys.}\ }\textbf
  {\bibinfo {volume} {71}},\ \bibinfo {pages} {463} (\bibinfo {year}
  {1999})}\BibitemShut {NoStop}%
\bibitem [{\citenamefont {Pitaevskii}\ and\ \citenamefont
  {Stringari}(2003)}]{Pitaevskii-Stringari-BEC}%
  \BibitemOpen
  \bibfield  {author} {\bibinfo {author} {\bibfnamefont {L.~P.}\ \bibnamefont
  {Pitaevskii}}\ and\ \bibinfo {author} {\bibfnamefont {S.}~\bibnamefont
  {Stringari}},\ }\href@noop {} {\emph {\bibinfo {title} {Bose-Einstein
  Condensation}}}\ (\bibinfo  {publisher} {Oxford University Press},\ \bibinfo
  {year} {2003})\BibitemShut {NoStop}%
\bibitem [{\citenamefont {Schley}\ \emph {et~al.}(2013)\citenamefont {Schley},
  \citenamefont {Berkovitz}, \citenamefont {Rinott}, \citenamefont {Shammass},
  \citenamefont {Blumkin},\ and\ \citenamefont {Steinhauer}}]{Schley-et-al}%
  \BibitemOpen
  \bibfield  {author} {\bibinfo {author} {\bibfnamefont {R.}~\bibnamefont
  {Schley}}, \bibinfo {author} {\bibfnamefont {A.}~\bibnamefont {Berkovitz}},
  \bibinfo {author} {\bibfnamefont {S.}~\bibnamefont {Rinott}}, \bibinfo
  {author} {\bibfnamefont {I.}~\bibnamefont {Shammass}}, \bibinfo {author}
  {\bibfnamefont {A.}~\bibnamefont {Blumkin}}, \ and\ \bibinfo {author}
  {\bibfnamefont {J.}~\bibnamefont {Steinhauer}},\ }\href {\doibase
  10.1103/PhysRevLett.111.055301} {\bibfield  {journal} {\bibinfo  {journal}
  {Phys. Rev. Lett.}\ }\textbf {\bibinfo {volume} {111}},\ \bibinfo {pages}
  {055301} (\bibinfo {year} {2013})}\BibitemShut {NoStop}%
\bibitem [{\citenamefont {Steinhauer}(2014)}]{Steinhauer-2014}%
  \BibitemOpen
  \bibfield  {author} {\bibinfo {author} {\bibfnamefont {J.}~\bibnamefont
  {Steinhauer}},\ }\href@noop {} {\bibfield  {journal} {\bibinfo  {journal}
  {Nat. Phys.}\ }\textbf {\bibinfo {volume} {10}},\ \bibinfo {pages} {864}
  (\bibinfo {year} {2014})}\BibitemShut {NoStop}%
\bibitem [{\citenamefont {Pethick}\ and\ \citenamefont
  {Smith}(2008)}]{Pethick-Smith-BEC}%
  \BibitemOpen
  \bibfield  {author} {\bibinfo {author} {\bibfnamefont {C.~J.}\ \bibnamefont
  {Pethick}}\ and\ \bibinfo {author} {\bibfnamefont {H.}~\bibnamefont
  {Smith}},\ }\href@noop {} {\emph {\bibinfo {title} {Bose-Einstein
  Condensation in Dilute Gases}}},\ \bibinfo {edition} {2nd}\ ed.\ (\bibinfo
  {publisher} {Cambridge University Press},\ \bibinfo {year}
  {2008})\BibitemShut {NoStop}%
\bibitem [{\citenamefont {Hong}\ \emph {et~al.}(1987)\citenamefont {Hong},
  \citenamefont {Ou},\ and\ \citenamefont {Mandel}}]{HOM}%
  \BibitemOpen
  \bibfield  {author} {\bibinfo {author} {\bibfnamefont {C.~K.}\ \bibnamefont
  {Hong}}, \bibinfo {author} {\bibfnamefont {Z.~Y.}\ \bibnamefont {Ou}}, \ and\
  \bibinfo {author} {\bibfnamefont {L.}~\bibnamefont {Mandel}},\ }\href
  {\doibase 10.1103/PhysRevLett.59.2044} {\bibfield  {journal} {\bibinfo
  {journal} {Phys. Rev. Lett.}\ }\textbf {\bibinfo {volume} {59}},\ \bibinfo
  {pages} {2044} (\bibinfo {year} {1987})}\BibitemShut {NoStop}%
\bibitem [{\citenamefont {Lopes}\ \emph {et~al.}(2015)\citenamefont {Lopes},
  \citenamefont {Imanaliev}, \citenamefont {Aspect}, \citenamefont {Cheneau},
  \citenamefont {Boiron},\ and\ \citenamefont {Westbrook}}]{Lopes-et-al}%
  \BibitemOpen
  \bibfield  {author} {\bibinfo {author} {\bibfnamefont {R.}~\bibnamefont
  {Lopes}}, \bibinfo {author} {\bibfnamefont {A.}~\bibnamefont {Imanaliev}},
  \bibinfo {author} {\bibfnamefont {A.}~\bibnamefont {Aspect}}, \bibinfo
  {author} {\bibfnamefont {M.}~\bibnamefont {Cheneau}}, \bibinfo {author}
  {\bibfnamefont {D.}~\bibnamefont {Boiron}}, \ and\ \bibinfo {author}
  {\bibfnamefont {C.~I.}\ \bibnamefont {Westbrook}},\ }\href
  {http://dx.doi.org/10.1038/nature14331} {\bibfield  {journal} {\bibinfo
  {journal} {Nature}\ }\textbf {\bibinfo {volume} {520}},\ \bibinfo {pages}
  {66} (\bibinfo {year} {2015})}\BibitemShut {NoStop}%
\bibitem [{\citenamefont {Unruh}(1981)}]{Unruh-1981}%
  \BibitemOpen
  \bibfield  {author} {\bibinfo {author} {\bibfnamefont {W.~G.}\ \bibnamefont
  {Unruh}},\ }\href {\doibase 10.1103/PhysRevLett.46.1351} {\bibfield
  {journal} {\bibinfo  {journal} {Phys. Rev. Lett.}\ }\textbf {\bibinfo
  {volume} {46}},\ \bibinfo {pages} {1351} (\bibinfo {year}
  {1981})}\BibitemShut {NoStop}%
\bibitem [{\citenamefont {Barcel{\'o}}\ \emph {et~al.}(2011)\citenamefont
  {Barcel{\'o}}, \citenamefont {Liberati},\ and\ \citenamefont
  {Visser}}]{AnalogueGravity-LivingReview}%
  \BibitemOpen
  \bibfield  {author} {\bibinfo {author} {\bibfnamefont {C.}~\bibnamefont
  {Barcel{\'o}}}, \bibinfo {author} {\bibfnamefont {S.}~\bibnamefont
  {Liberati}}, \ and\ \bibinfo {author} {\bibfnamefont {M.}~\bibnamefont
  {Visser}},\ }\href {\doibase 10.12942/lrr-2011-3} {\bibfield  {journal}
  {\bibinfo  {journal} {Living Reviews in Relativity}\ }\textbf {\bibinfo
  {volume} {14}},\ \bibinfo {pages} {3} (\bibinfo {year} {2011})}\BibitemShut
  {NoStop}%
\bibitem [{\citenamefont {Macher}\ and\ \citenamefont
  {Parentani}(2009)}]{Macher-Parentani-BEC}%
  \BibitemOpen
  \bibfield  {author} {\bibinfo {author} {\bibfnamefont {J.}~\bibnamefont
  {Macher}}\ and\ \bibinfo {author} {\bibfnamefont {R.}~\bibnamefont
  {Parentani}},\ }\href {\doibase 10.1103/PhysRevA.80.043601} {\bibfield
  {journal} {\bibinfo  {journal} {Phys. Rev. A}\ }\textbf {\bibinfo {volume}
  {80}},\ \bibinfo {pages} {043601} (\bibinfo {year} {2009})}\BibitemShut
  {NoStop}%
\bibitem [{\citenamefont {Michel}\ and\ \citenamefont
  {Parentani}(2015)}]{Michel-Parentani-2015}%
  \BibitemOpen
  \bibfield  {author} {\bibinfo {author} {\bibfnamefont {F.}~\bibnamefont
  {Michel}}\ and\ \bibinfo {author} {\bibfnamefont {R.}~\bibnamefont
  {Parentani}},\ }\href {\doibase 10.1103/PhysRevA.91.053603} {\bibfield
  {journal} {\bibinfo  {journal} {Phys. Rev. A}\ }\textbf {\bibinfo {volume}
  {91}},\ \bibinfo {pages} {053603} (\bibinfo {year} {2015})}\BibitemShut
  {NoStop}%
\bibitem [{\citenamefont {Michel}\ \emph
  {et~al.}(2016{\natexlab{a}})\citenamefont {Michel}, \citenamefont
  {Parentani},\ and\ \citenamefont {Zegers}}]{Michel-Parentani-Zegers-2016}%
  \BibitemOpen
  \bibfield  {author} {\bibinfo {author} {\bibfnamefont {F.}~\bibnamefont
  {Michel}}, \bibinfo {author} {\bibfnamefont {R.}~\bibnamefont {Parentani}}, \
  and\ \bibinfo {author} {\bibfnamefont {R.}~\bibnamefont {Zegers}},\ }\href
  {\doibase 10.1103/PhysRevD.93.065039} {\bibfield  {journal} {\bibinfo
  {journal} {Phys. Rev. D}\ }\textbf {\bibinfo {volume} {93}},\ \bibinfo
  {pages} {065039} (\bibinfo {year} {2016}{\natexlab{a}})}\BibitemShut
  {NoStop}%
\bibitem [{\citenamefont {Michel}\ \emph
  {et~al.}(2016{\natexlab{b}})\citenamefont {Michel}, \citenamefont
  {Coupechoux},\ and\ \citenamefont
  {Parentani}}]{Michel-Coupechoux-Parentani-2016}%
  \BibitemOpen
  \bibfield  {author} {\bibinfo {author} {\bibfnamefont {F.}~\bibnamefont
  {Michel}}, \bibinfo {author} {\bibfnamefont {J.-F.}\ \bibnamefont
  {Coupechoux}}, \ and\ \bibinfo {author} {\bibfnamefont {R.}~\bibnamefont
  {Parentani}},\ }\href {\doibase 10.1103/PhysRevD.94.084027} {\bibfield
  {journal} {\bibinfo  {journal} {Phys. Rev. D}\ }\textbf {\bibinfo {volume}
  {94}},\ \bibinfo {pages} {084027} (\bibinfo {year}
  {2016}{\natexlab{b}})}\BibitemShut {NoStop}%
\bibitem [{\citenamefont {Boiron}\ \emph {et~al.}(2015)\citenamefont {Boiron},
  \citenamefont {Fabbri}, \citenamefont {Larr\'e}, \citenamefont {Pavloff},
  \citenamefont {Westbrook},\ and\ \citenamefont {Zi\ifmmode~\acute{n}\else
  \'{n}\fi{}}}]{Boiron-et-al}%
  \BibitemOpen
  \bibfield  {author} {\bibinfo {author} {\bibfnamefont {D.}~\bibnamefont
  {Boiron}}, \bibinfo {author} {\bibfnamefont {A.}~\bibnamefont {Fabbri}},
  \bibinfo {author} {\bibfnamefont {P.-E.}\ \bibnamefont {Larr\'e}}, \bibinfo
  {author} {\bibfnamefont {N.}~\bibnamefont {Pavloff}}, \bibinfo {author}
  {\bibfnamefont {C.~I.}\ \bibnamefont {Westbrook}}, \ and\ \bibinfo {author}
  {\bibfnamefont {P.}~\bibnamefont {Zi\ifmmode~\acute{n}\else \'{n}\fi{}}},\
  }\href {\doibase 10.1103/PhysRevLett.115.025301} {\bibfield  {journal}
  {\bibinfo  {journal} {Phys. Rev. Lett.}\ }\textbf {\bibinfo {volume} {115}},\
  \bibinfo {pages} {025301} (\bibinfo {year} {2015})}\BibitemShut {NoStop}%
\bibitem [{\citenamefont {Euv\'e}\ \emph {et~al.}(2016)\citenamefont {Euv\'e},
  \citenamefont {Michel}, \citenamefont {Parentani}, \citenamefont {Philbin},\
  and\ \citenamefont {Rousseaux}}]{Euve-et-al-2016}%
  \BibitemOpen
  \bibfield  {author} {\bibinfo {author} {\bibfnamefont {L.-P.}\ \bibnamefont
  {Euv\'e}}, \bibinfo {author} {\bibfnamefont {F.}~\bibnamefont {Michel}},
  \bibinfo {author} {\bibfnamefont {R.}~\bibnamefont {Parentani}}, \bibinfo
  {author} {\bibfnamefont {T.~G.}\ \bibnamefont {Philbin}}, \ and\ \bibinfo
  {author} {\bibfnamefont {G.}~\bibnamefont {Rousseaux}},\ }\href {\doibase
  10.1103/PhysRevLett.117.121301} {\bibfield  {journal} {\bibinfo  {journal}
  {Phys. Rev. Lett.}\ }\textbf {\bibinfo {volume} {117}},\ \bibinfo {pages}
  {121301} (\bibinfo {year} {2016})}\BibitemShut {NoStop}%
\bibitem [{\citenamefont {Balbinot}\ \emph {et~al.}(2008)\citenamefont
  {Balbinot}, \citenamefont {Fabbri}, \citenamefont {Fagnocchi}, \citenamefont
  {Recati},\ and\ \citenamefont {Carusotto}}]{Balbinot-et-al-2008}%
  \BibitemOpen
  \bibfield  {author} {\bibinfo {author} {\bibfnamefont {R.}~\bibnamefont
  {Balbinot}}, \bibinfo {author} {\bibfnamefont {A.}~\bibnamefont {Fabbri}},
  \bibinfo {author} {\bibfnamefont {S.}~\bibnamefont {Fagnocchi}}, \bibinfo
  {author} {\bibfnamefont {A.}~\bibnamefont {Recati}}, \ and\ \bibinfo {author}
  {\bibfnamefont {I.}~\bibnamefont {Carusotto}},\ }\href {\doibase
  10.1103/PhysRevA.78.021603} {\bibfield  {journal} {\bibinfo  {journal} {Phys.
  Rev. A}\ }\textbf {\bibinfo {volume} {78}},\ \bibinfo {pages} {021603}
  (\bibinfo {year} {2008})}\BibitemShut {NoStop}%
\bibitem [{\citenamefont {Carusotto}\ \emph {et~al.}(2008)\citenamefont
  {Carusotto}, \citenamefont {Fagnocchi}, \citenamefont {Recati}, \citenamefont
  {Balbinot},\ and\ \citenamefont {Fabbri}}]{Carusotto-BH}%
  \BibitemOpen
  \bibfield  {author} {\bibinfo {author} {\bibfnamefont {I.}~\bibnamefont
  {Carusotto}}, \bibinfo {author} {\bibfnamefont {S.}~\bibnamefont
  {Fagnocchi}}, \bibinfo {author} {\bibfnamefont {A.}~\bibnamefont {Recati}},
  \bibinfo {author} {\bibfnamefont {R.}~\bibnamefont {Balbinot}}, \ and\
  \bibinfo {author} {\bibfnamefont {A.}~\bibnamefont {Fabbri}},\ }\href
  {http://stacks.iop.org/1367-2630/10/i=10/a=103001} {\bibfield  {journal}
  {\bibinfo  {journal} {New Journal of Physics}\ }\textbf {\bibinfo {volume}
  {10}},\ \bibinfo {pages} {103001} (\bibinfo {year} {2008})}\BibitemShut
  {NoStop}%
\bibitem [{\citenamefont {Einstein}\ \emph {et~al.}(1935)\citenamefont
  {Einstein}, \citenamefont {Podolsky},\ and\ \citenamefont {Rosen}}]{EPR}%
  \BibitemOpen
  \bibfield  {author} {\bibinfo {author} {\bibfnamefont {A.}~\bibnamefont
  {Einstein}}, \bibinfo {author} {\bibfnamefont {B.}~\bibnamefont {Podolsky}},
  \ and\ \bibinfo {author} {\bibfnamefont {N.}~\bibnamefont {Rosen}},\ }\href
  {\doibase 10.1103/PhysRev.47.777} {\bibfield  {journal} {\bibinfo  {journal}
  {Phys. Rev.}\ }\textbf {\bibinfo {volume} {47}},\ \bibinfo {pages} {777}
  (\bibinfo {year} {1935})}\BibitemShut {NoStop}%
\bibitem [{\citenamefont {Schr\"{o}dinger}(1935)}]{Schroedinger-1935}%
  \BibitemOpen
  \bibfield  {author} {\bibinfo {author} {\bibfnamefont {E.}~\bibnamefont
  {Schr\"{o}dinger}},\ }\href {\doibase 10.1017/S0305004100013554} {\bibfield
  {journal} {\bibinfo  {journal} {Proc. Camb. Phil. Soc.}\ }\textbf {\bibinfo
  {volume} {31}},\ \bibinfo {pages} {555} (\bibinfo {year} {1935})}\BibitemShut
  {NoStop}%
\bibitem [{\citenamefont {Reid}(1989)}]{Reid-1989}%
  \BibitemOpen
  \bibfield  {author} {\bibinfo {author} {\bibfnamefont {M.~D.}\ \bibnamefont
  {Reid}},\ }\href {\doibase 10.1103/PhysRevA.40.913} {\bibfield  {journal}
  {\bibinfo  {journal} {Phys. Rev. A}\ }\textbf {\bibinfo {volume} {40}},\
  \bibinfo {pages} {913} (\bibinfo {year} {1989})}\BibitemShut {NoStop}%
\bibitem [{\citenamefont {Peise}\ \emph {et~al.}(2015)\citenamefont {Peise},
  \citenamefont {Kruse}, \citenamefont {Lange}, \citenamefont {L{\"u}cke},
  \citenamefont {Pezz{\`e}}, \citenamefont {Arlt}, \citenamefont {Ertmer},
  \citenamefont {Hammerer}, \citenamefont {Santos}, \citenamefont {Smerzi},\
  and\ \citenamefont {Klempt}}]{Peise-et-al-2015}%
  \BibitemOpen
  \bibfield  {author} {\bibinfo {author} {\bibfnamefont {J.}~\bibnamefont
  {Peise}}, \bibinfo {author} {\bibfnamefont {I.}~\bibnamefont {Kruse}},
  \bibinfo {author} {\bibfnamefont {K.}~\bibnamefont {Lange}}, \bibinfo
  {author} {\bibfnamefont {B.}~\bibnamefont {L{\"u}cke}}, \bibinfo {author}
  {\bibfnamefont {L.}~\bibnamefont {Pezz{\`e}}}, \bibinfo {author}
  {\bibfnamefont {J.}~\bibnamefont {Arlt}}, \bibinfo {author} {\bibfnamefont
  {W.}~\bibnamefont {Ertmer}}, \bibinfo {author} {\bibfnamefont
  {K.}~\bibnamefont {Hammerer}}, \bibinfo {author} {\bibfnamefont
  {L.}~\bibnamefont {Santos}}, \bibinfo {author} {\bibfnamefont
  {A.}~\bibnamefont {Smerzi}}, \ and\ \bibinfo {author} {\bibfnamefont
  {C.}~\bibnamefont {Klempt}},\ }\href {http://dx.doi.org/10.1038/ncomms9984}
  {\bibfield  {journal} {\bibinfo  {journal} {Nature Communications}\ }\textbf
  {\bibinfo {volume} {6}},\ \bibinfo {pages} {8984 EP } (\bibinfo {year}
  {2015})}\BibitemShut {NoStop}%
\bibitem [{\citenamefont {Ross}(2012)}]{Ross}%
  \BibitemOpen
  \bibfield  {author} {\bibinfo {author} {\bibfnamefont {S.}~\bibnamefont
  {Ross}},\ }\href@noop {} {\emph {\bibinfo {title} {A First Course in
  Probability}}},\ \bibinfo {edition} {9th}\ ed.\ (\bibinfo  {publisher}
  {Pearson},\ \bibinfo {year} {2012})\BibitemShut {NoStop}%
\end{thebibliography}%

\end{document}